\documentclass[11pt,a4paper]{article}

\usepackage{amsmath,amssymb}
\usepackage[margin=1in]{geometry}
\usepackage{graphicx}
\usepackage{hyperref}

\begin{document}

\title{Dark matter from the quadratic spinor Lagrangian. I.\\
Geometric mass for a gravitationally produced spin-$1/2$ fermion}

\author{Roh-Suan Tung\\[4pt]
\small\it Department of Physics and Center for Theoretical Physics, \\ 
\small\it Chung Yuan Christian University, Taoyuan 320, Taiwan\\
\small\it Institute of Advanced Studies, Nanyang Technological University, 639673, Singapore\\[2pt]
\small\texttt{rstung@cycu.edu.tw}}

\date{ }

\maketitle

\begin{abstract}
The gravitational-wave induced freeze-in of Maleknejad and Kopp (2026)
produces dark fermions from a stochastic gravitational-wave background,
but requires them to acquire mass by separate means. We develop the
Quadratic Spinor Lagrangian (QSL) formulation of general relativity,
extended to Einstein--Cartan, as a framework that supplies this mass
geometrically. The spinor 1-form built from a single Dirac field is
purely spin-$1/2$---its gamma-traceless (spin-$3/2$) part vanishes
identically---so the propagating excitation is a Dirac fermion, the same
content as the produced Weyl fermion. A cosmological spinor condensate
sources a vectorial trace torsion $K\propto\dot\chi/\chi$, and an explicit
Clifford reduction shows that this torsion gives the fermion a pure Dirac
mass $M_{\rm eff}=(1/\sqrt6)\,|\dot\chi/\chi|$, with no pseudoscalar or
cross terms. The mass is not a free parameter but is locked to the Hubble
rate at production, $M_{\rm eff}\simeq(c_\chi/\sqrt6)H_*$, making the relic
abundance a function of essentially the single scale $H_*$
($\Omega h^2\propto H_*^{5/2}$) and supplying the mass the parent
mechanism must postulate. Whether promoting the spinor 1-form to an
independent field yields a propagating spin-$3/2$ candidate is a distinct
dynamical question; Paper~II~\cite{TungNoGo} shows that it does
not---the QSL channels all propagation into the gravitational sector---so
the composite spin-$1/2$ Dirac fermion is the unique QSL dark-matter
candidate. We discuss the resulting dark-matter phenomenology and its link
to asymptotically free scalar-field cosmology.
\end{abstract}


\section{Introduction}
\label{sec:intro}

Maleknejad and Kopp~\cite{MaleknejadKopp2026,MaleknejadKopp2025companion} recently
demonstrated that stochastic gravitational wave (GW) backgrounds break the conformal
symmetry of massless Weyl fermions in the early universe. While the expansion of a conformally flat
Friedmann--Lema\^itre--Robertson--Walker (FLRW) universe preserves fermion conformal
symmetry---the energy density dilutes as $a^{-4}$ and no net particle production
occurs---the presence of GWs introduces transverse-traceless perturbations that
cannot be absorbed by a conformal rescaling. This enables an efficient gravitational
freeze-in mechanism for dark matter (DM) production.

To complete this model, the authors require two external conditions: (i) a stochastic
GW background sourced by specific cosmological events (phase transitions, preheating,
cosmic strings), and (ii) a late-time mass-generation mechanism to render the
produced fermions non-relativistic.

In this paper, we develop the Quadratic Spinor Lagrangian (QSL) for general
relativity~\cite{NesterTung1995,TungJacobson1995} as a comprehensive framework for
understanding and extending this mechanism. The paper is organized as follows.

In section~\ref{sec:QSL} we review the QSL formalism and the established results
that are relevant: the spinor-curvature identity, the metric as a spinor bilinear,
the Yang--Mills analogy, and the super-$SL(2,\mathbb{C})$ gauge structure.
Section~\ref{sec:conformal} uses the spinor-curvature identity to provide a
structural explanation for conformal symmetry breaking by GWs.
Section~\ref{sec:vertices} derives the fermion--graviton interaction vertices
covariantly within the QSL perturbation theory.

The remaining sections develop new results.
Section~\ref{sec:EinsteinCartan} extends the QSL to the Einstein--Cartan setting,
deriving the trace torsion generated by a cosmological spinor condensate.
Section~\ref{sec:spin32} analyzes the spin content of the spinor 1-form via
Lorentz representation theory, showing that the composite field is purely
spin-$1/2$ while a spin-$3/2$ sector opens only upon promoting it to an
independent field (left to future work).
Section~\ref{sec:mass} derives the Dirac mass that trace torsion confers on
the propagating spin-$1/2$ fermion.
Section~\ref{sec:DM} discusses dark matter phenomenology, and
section~\ref{sec:cosmology} connects the framework to asymptotically free
scalar field cosmology.
Section~\ref{sec:spin32ext} gathers the analyses bearing on the
independent spin-$3/2$ extension---the Rarita--Schwinger comparison,
causality, and the cosmological constraint reduction---which we leave to
future work.
We conclude in section~\ref{sec:discussion}.

A single structural feature unifies the new results. The torsion-generated
mass is not a free parameter but is locked to the expansion rate at
production: $M_{\rm eff}=(1/\sqrt6)|\dot\chi/\chi|\simeq(c_\chi/\sqrt6)H_*$,
where the Clifford coefficient $1/\sqrt6$ is exact while
$c_\chi\equiv\dot\chi/(\chi H)=O(1)$ is fixed by the condensate dynamics. For
the spin-$1/2$ Dirac fermion that the QSL actually propagates, this turns
the relic abundance into a one-parameter prediction
(section~\ref{sec:DM}) and supplies geometrically the mass that the
parent mechanism postulates. The same locking, applied to the future
spin-$3/2$ extension, keeps the helicity-$\tfrac12$ sound speed strictly
positive (section~\ref{sec:longitudinal})---a structural reason to expect
that extension to be causally well behaved, which together with the
metric-cone propagation of section~\ref{sec:characteristic} motivates its
development. A concise account of the central result---the geometric
Dirac mass and its one-parameter relic prediction---is given in the
companion Letter~\cite{TungLetter}.

\textit{Notation and conventions.}
We use the ``Clifform'' notation of~\cite{DimakisMullerHoissen1991}.
We use the mostly-plus signature $g_{ab}=\mathrm{diag}(-,+,+,+)$ throughout,
so that $\gamma_0^2=-1$ and $\gamma_i^2=+1$. Dirac matrix conventions:
$\gamma_{(a}\gamma_{b)} = g_{ab}$ (consistent with the signature, since
$\gamma_{(0}\gamma_{0)}=g_{00}=-1$),
$\gamma_{ab} := \gamma_{[a}\gamma_{b]}$,
$\gamma_5 := \gamma_0\gamma_1\gamma_2\gamma_3$ (equivalently
$\gamma_5=i\gamma^0\gamma^1\gamma^2\gamma^3$, as used in
appendix~\ref{app:clifford}); the resulting factor of $i$
in $\gamma_5$ bilinears is the universal QSL reality convention (it
multiplies kinetic and mass terms alike and cancels in physical ratios).
A timelike vector then has negative square, e.g.\ the trace torsion
$K^aK_a=-K^2<0$. The wedge product between forms is often suppressed.
$D$ denotes the covariant exterior derivative, $\mathring{D}$ the
torsion-free (Levi-Civita) covariant exterior derivative.
Greek indices $\mu,\nu,\ldots$ are spacetime coordinate indices;
Latin indices $a,b,\ldots$ are orthonormal frame (Lorentz) indices;
upper-case Latin indices $A,B,\ldots$ are two-component spinor indices.

\section{The Quadratic Spinor Lagrangian}
\label{sec:QSL}

\subsection{Action and spinor-curvature identity}
\label{sec:QSL:action}

Let $\vartheta^a = e^a_\mu\,dx^\mu$ be the tetrad 1-form and $\psi$ a
Dirac spinor field normalized according to $\bar\psi\psi = 1$,
$\bar\psi\gamma_5\psi = 0$. Define the Clifford-algebra-valued 1-form
\begin{equation}
    \Psi := \vartheta^a\gamma_a\psi \,.
    \label{eq:Psi_def}
\end{equation}
The Quadratic Spinor Lagrangian is the 4-form~\cite{NesterTung1995,TungNester1999}
\begin{equation}
    \mathcal{L}_{\rm QSL} = 2\,D\Psi\,\gamma_5\,D\Psi \,,
    \label{eq:QSL}
\end{equation}
where $D$ is the covariant exterior derivative constructed from the
spin connection $\omega^{ab}$, acting on both the spinor and frame indices of $\Psi$.

The spinor-curvature identity~\cite{NesterTungZhytnikov1994} relates the QSL to
the Einstein--Hilbert Lagrangian:
\begin{equation}
    2\,D\Psi\,\gamma_5\,D\Psi
    \equiv -\bar\psi\psi\;R\,{*1}
    + d\bigl[(D\Psi)\gamma_5\Psi + \Psi\gamma_5(D\Psi)\bigr] \,.
    \label{eq:spinor_curvature_identity}
\end{equation}
For normalized $\psi$, the QSL differs from the Hilbert scalar curvature
Lagrangian only by an exact differential (boundary term). Equivalently,
\begin{equation}
    2\,D\Psi\,\gamma_5\,D\Psi
    \equiv 2\,\Psi\,\Omega\,\gamma_5\,\Psi
    + d\bigl[(D\Psi)\gamma_5\Psi + \Psi\gamma_5(D\Psi)\bigr] \,,
    \label{eq:identity_curvature}
\end{equation}
where $\Omega = \frac{1}{4}\Omega_{ab}\gamma^{ab} = d\omega + \omega\wedge\omega$
is the Clifford-algebra-valued curvature 2-form. The boundary term is the
same exact differential as in~\eqref{eq:spinor_curvature_identity}: applying
the graded Leibniz rule to the $3$-form
$\Theta=(D\Psi)\gamma_5\Psi+\Psi\gamma_5(D\Psi)$, with $D\gamma_5=0$ and
$D^2\Psi=\Omega\Psi$, gives
$d\Theta=2\,D\Psi\,\gamma_5\,D\Psi-2\,\Psi\,\Omega\,\gamma_5\,\Psi$ (using
$\gamma_5\Omega=\Omega\gamma_5$ and the $2$-form symmetry
$(\Omega\Psi)\gamma_5\Psi=-\Psi\,\Omega\,\gamma_5\Psi$), which is
precisely~\eqref{eq:identity_curvature}. The two forms of the identity share
this boundary term because their bulk densities are algebraically equal,
$2\,\Psi\,\Omega\,\gamma_5\,\Psi=-\bar\psi\psi\,R\,{*1}$ for normalized $\psi$,
the first Bianchi identity $\Omega_{ab}\wedge\vartheta^b=0$ collapsing the
curvature onto its Ricci trace (appendix~\ref{app:curvature}).

The field equations follow by independent variation of the spinor
1-form $\Psi$ (equivalently $\psi$) and the spin connection
$\omega^{ab}$. Three properties of the formalism are used repeatedly:
$D$ is a graded derivation,
$D(\phi\,\eta)=D\phi\,\eta+(-1)^p\phi\,D\eta$ for a $p$-form
$\phi$; the chirality element is covariantly constant, $D\gamma_5=0$,
since $\gamma_5$ commutes with the connection generators $\gamma^{ab}$;
and the $\gamma_5$ bilinear is symmetric on even forms, so that for two
Clifford-algebra-valued 2-forms $A,B$ one has $A\gamma_5 B=B\gamma_5 A$.

\paragraph{Variation of $\Psi$.}
With $\omega$ held fixed, $\delta(D\Psi)=D(\delta\Psi)$, and the symmetry
of the bilinear makes the two resulting terms equal:
\begin{equation}
    \delta_\Psi\mathcal{L}_{\rm QSL}
    = 2\,D(\delta\Psi)\,\gamma_5\,D\Psi + 2\,D\Psi\,\gamma_5\,D(\delta\Psi)
    = 4\,D(\delta\Psi)\,\gamma_5\,D\Psi \,.
\end{equation}
Integrating by parts, with $\delta\Psi$ a 1-form and $D\gamma_5=0$,
\begin{equation}
    D\!\left(\delta\Psi\,\gamma_5\,D\Psi\right)
    = D(\delta\Psi)\,\gamma_5\,D\Psi - \delta\Psi\,\gamma_5\,D^2\Psi \,,
\end{equation}
so that
\begin{equation}
    \delta_\Psi S
    = 4\oint \delta\Psi\,\gamma_5\,D\Psi
    + 4\int \delta\Psi\,\gamma_5\,D^2\Psi \,.
\end{equation}
The $O(1/r^4)$ fall-off of the QSL annihilates the boundary term, and
stationarity for arbitrary $\delta\Psi$ yields the Einstein equation.

\paragraph{Variation of $\omega^{ab}$.}
The connection enters only through $D$, and does so algebraically:
$\delta_\omega(D\Psi)=\delta\omega\,\Psi$ with
$\delta\omega=\tfrac14\delta\omega_{ab}\gamma^{ab}$ a 1-form. Hence
\begin{equation}
    \delta_\omega\mathcal{L}_{\rm QSL}
    = 2\,(\delta\omega\,\Psi)\,\gamma_5\,D\Psi
    + 2\,D\Psi\,\gamma_5\,(\delta\omega\,\Psi) \,.
\end{equation}
Equivalently, using the identity form $2\Psi\Omega\gamma_5\Psi$ with
$\delta\Omega=D\delta\omega$ and integrating by parts, the derivative on
$\delta\omega$ is removed, leaving an expression algebraic in
$\delta\omega_{ab}$. Stationarity then gives the covariantly conserved
spin current $D(\bar\Psi\gamma_{ab}\gamma_5\Psi)=0$; since
$\bar\Psi\gamma_{ab}\gamma_5\Psi$ is a tetrad bilinear ($\sim e\wedge e$),
this is the QSL analogue of the Palatini relation
$D(\vartheta^a\wedge\vartheta^b)=0$ and is algebraic in the torsion,
forcing $T^a=0$ in vacuum.

Collecting both variations, the field equations are
\begin{align}
    D^2\Psi &= 0 \qquad\text{(Einstein equation)}\,,
    \label{eq:Einstein}\\
    D(\bar\Psi\gamma_{ab}\gamma_5\Psi) &= 0 \qquad\text{(torsion-free condition)}\,.
    \label{eq:torsion_free}
\end{align}
Equation~\eqref{eq:Einstein}, written with $D^2\Psi=\Omega\,\Psi$ and
expanded using the first Bianchi identity
$\Omega_{ab}\wedge\vartheta^b=0$ to collapse the curvature onto its
trace, yields the ``Clifform'' transcription of the vacuum Einstein
equation:
$G^a{}_b\,\gamma_5\gamma^b\,{*\vartheta_a}\,\psi = 0$~\cite{DimakisMullerHoissen1991}.

A key feature of the QSL is its asymptotic fall-off: $\mathcal{L}_{\rm QSL}$
is $O(1/r^4)$ rather than the $O(1/r^3)$ of the Einstein--Hilbert Lagrangian.
This guarantees a finite action and a well-defined, fully 4-covariant
Hamiltonian~\cite{NesterTung1995}.

\subsection{Spinor 1-forms and the metric}
\label{sec:QSL:metric}

An equivalent formulation~\cite{TungJacobson1995} takes as fundamental
variables two $SL(2,\mathbb{C})$ spinor 1-forms $\phi^A$, $\chi^A$,
related to the tetrad by $\phi^A = \theta^{A0'}$, $\chi^A = \theta^{A1'}$.
The chiral action is
\begin{equation}
    S_2[\chi^A,\phi^A,\omega^{AB}]
    = 4i\int D\phi^A\wedge D\chi_A \,,
    \label{eq:chiral_action}
\end{equation}
with field equations $D^2\phi^A = 0$, $D^2\chi^A = 0$ (Einstein equation),
and $D(\chi^{(A}\wedge\phi^{B)}) = 0$ (torsion vanishes).

Combining into a Dirac spinor 1-form $\Psi = (\phi^A, \chi^{A'})$,
the metric is a spinor bilinear~\cite{TungJacobson1995}:
\begin{equation}
    g = \Psi\otimes_S\Psi \,,\qquad
    g_{\mu\nu} = \bar\Psi_{(\mu}\Psi_{\nu)} \,,
    \label{eq:metric_from_spinor}
\end{equation}
where $\otimes_S$ denotes the symmetrized tensor product.
A GW perturbation $\delta g_{\mu\nu}$ therefore corresponds to a
perturbation $\delta\Psi$ of the underlying spinor 1-form:
\begin{equation}
    \delta g_{\mu\nu} = \bar\Psi_{(\mu}\,\delta\Psi_{\nu)}
    + \delta\bar\Psi_{(\mu}\,\Psi_{\nu)} \,.
    \label{eq:metric_perturbation}
\end{equation}

\subsection{Yang--Mills analogy}
\label{sec:QSL:YM}

The QSL carries a close formal analogy with Yang--Mills gauge
theory~\cite{TungNester2000}. Recall the Yang--Mills data: a gauge group
with Lie algebra generators $T_I$ ($I=1,\dots,\dim\mathfrak g$) obeying
$[T_I,T_J]=f_{IJ}{}^{K}T_K$, a connection 1-form
$A = A^I{}_\mu\,T_I\,dx^\mu$ (the index $I$ labelling the algebra
direction), and its curvature 2-form
\begin{equation}
    F = dA + A\wedge A \,,
\end{equation}
in terms of which the action is $\tfrac12\!\int\!\mathrm{tr}(F\wedge{*F})$
and the field equation is $D{*F}=0$, with $D$ the gauge-covariant
exterior derivative and $*$ the Hodge dual.

The QSL has the same structure with the Lie-algebra index traded for the
Clifford (spinor--frame) structure. The spinor 1-form $\Psi=\vartheta^a
\gamma_a\psi$ plays the role of the potential, and the gravitational
field strength is its covariant derivative,
\begin{equation}
    F_G := D\Psi \,,
    \label{eq:field_strength}
\end{equation}
so that $\mathcal{L}_{\rm QSL}=2\,F_G\,\gamma_5\,F_G$ and the Einstein
equation takes the Yang--Mills form $D\gamma_5 F_G=0$. The chirality
element $\gamma_5$ is the QSL counterpart of the Hodge dual $*$: it is
the metric-independent involution that builds the quadratic invariant.
The correspondence is summarized below.

\begin{center}
\begin{tabular}{lcc}
\hline\hline
 & Yang--Mills & Quadratic spinor GR \\ \hline
Potential & $A = A^I{}_\mu\,T_I\,dx^\mu$ & $\Psi = e^a{}_\mu\,\gamma_a\psi\,dx^\mu$ \\
Field strength & $F = dA+A\wedge A$ & $F_G = D\Psi$ \\
Quadratic involution & Hodge dual ${*}$ & chirality $\gamma_5$ \\
Lagrangian & $\tfrac12\,\mathrm{tr}\,F\wedge{*F}$ & $2\,F_G\,\gamma_5\,F_G$ \\
Field equation & $D{*F} = 0$ & $D\gamma_5\,F_G = 0$ \\
\hline\hline
\end{tabular}
\end{center}

\noindent
The $D\gamma_5 F_G=0$ form, gauge-degenerate exactly as the Yang--Mills
equation is, is what we use in the causal analysis of
section~\ref{sec:characteristic}.

\subsection{Super-$SL(2,\mathbb{C})$ gauge structure}
\label{sec:QSL:super}

The QSL further admits a super-$SL(2,\mathbb{C})$
extension~\cite{Tung2000} that assembles the spin connection
$\omega^{AB}$ and the spinor 1-form $\phi^A$ into a single graded
connection whose algebra closes with $\{Q_A,Q_B\}=2M_{AB}$: the
anticommutator of two fermionic (spinorial) generators is a bosonic
(Lorentz) one, so gravitational and spinor excitations are algebraically
intertwined. This is the structural counterpart of the production
mechanism of section~\ref{sec:DM} and foreshadows the supergravity-like
character of the Einstein--Cartan completion
(section~\ref{sec:EinsteinCartan}).

\section{Conformal symmetry breaking by gravitational waves}
\label{sec:conformal}

The spinor-curvature identity~\eqref{eq:identity_curvature} provides a
transparent explanation for the central result of
Ref.~\cite{MaleknejadKopp2026}.

\subsection{FLRW background: conformal flatness}

In a spatially flat FLRW universe with scale factor $a(\tau)$, the metric
is conformally flat: $ds^2 = a^2(\tau)(-d\tau^2 + \delta_{ij}dx^idx^j)$.
The Weyl tensor vanishes identically, and the curvature 2-form $\Omega_{ab}$
is purely Ricci-type. The standard conformal rescaling
$\psi \to a^{3/2}\psi$ absorbs all curvature effects, and the canonically
normalized fermion field evolves freely. The fermion energy density dilutes
as $a^{-4}$, and no net particle production occurs.

In the language of the spinor-curvature identity~\eqref{eq:identity_curvature},
the curvature 2-form $\Omega$ in a conformally flat background contributes
only through the conformal part, which can be removed by the rescaling. The
``source term'' $\Psi\,\Omega\,\gamma_5\,\Psi$ is effectively zero for conformally
coupled fermions.

\subsection{Gravitational wave perturbations: Weyl curvature}

Now consider the metric perturbed by GWs:
$ds^2 = a^2(\tau)\bigl[-d\tau^2 + (\delta_{ij} + h_{ij})dx^idx^j\bigr]$,
where $h_{ij}$ is transverse-traceless. The curvature 2-form acquires
additional components:
\begin{equation}
    \Omega_{ab} = \mathring{\Omega}_{ab}^{(\rm FLRW)}
    + \delta\Omega_{ab}(h) + O(h^2) \,,
    \label{eq:curvature_decomposition}
\end{equation}
where $\delta\Omega_{ab}(h)$ carries transverse-traceless (Weyl-type) components.
These cannot be removed by any conformal rescaling of the spinor field.

Substituting into the identity~\eqref{eq:identity_curvature}:
\begin{equation}
    2\,\Psi\,\Omega\,\gamma_5\,\Psi
    = \underbrace{2\,\Psi\,\mathring{\Omega}^{(\rm FLRW)}\gamma_5\,\Psi}_{\text{absorbed by rescaling}}
    + \underbrace{2\,\Psi\,\delta\Omega(h)\,\gamma_5\,\Psi}_{\text{NOT absorbable}} + \cdots
    \label{eq:conformal_breaking}
\end{equation}
The second term is the structural origin of conformal symmetry breaking.
It acts as a source for spinor excitations precisely when the curvature
deviates from its conformally flat form; the explicit linearized
coupling $\delta\omega^{ab}(h)$ is derived in section~\ref{sec:vertices}.

\section{Covariant fermion--graviton vertices}
\label{sec:vertices}

In Ref.~\cite{MaleknejadKopp2026}, the cubic and quartic fermion--graviton
interaction vertices [their Eqs.~(4)--(5)] are derived by expanding the
spinor covariant derivative around the FLRW background in the
transverse-traceless gauge. Here we show that the QSL provides a covariant
derivation of the same vertices.

\subsection{Perturbation of the spin connection}

Write the full spin connection as
$\omega^{ab} = \mathring{\omega}^{ab} + \delta\omega^{ab}(h)$,
where $\mathring\omega^{ab}$ is the FLRW background connection and
$\delta\omega^{ab}(h)$ encodes the GW perturbation. Then
\begin{equation}
    D\Psi = \mathring{D}\Psi + \delta\omega\,\Psi \,,
    \label{eq:D_expansion}
\end{equation}
and the QSL Lagrangian expands as
\begin{equation}
    \mathcal{L}_{\rm QSL}
    = 2\,\mathring{D}\Psi\,\gamma_5\,\mathring{D}\Psi
    + 2\bigl(\delta\omega\,\Psi\,\gamma_5\,\mathring{D}\Psi
      + \mathring{D}\Psi\,\gamma_5\,\delta\omega\,\Psi\bigr)
    + 2\,\delta\omega\,\Psi\,\gamma_5\,\delta\omega\,\Psi \,.
    \label{eq:QSL_expansion}
\end{equation}

\subsection{Identification with Maleknejad--Kopp vertices}

For a transverse-traceless perturbation $g_{ij}=a^2(\delta_{ij}+h_{ij})$
($\partial^i h_{ij}=0$, $h^i{}_i=0$) the perturbed tetrad in conformal
time is $e^0=a\,d\eta$, $e^i=a(\delta^i_j+\tfrac12 h^i_j)\,dx^j$, and the
torsion-free condition $de^a+\omega^a{}_b\wedge e^b=0$ gives the
linearized spin-connection components
\begin{equation}
    \delta\omega^{0i} = \tfrac12 h'_{ij}\,dx^j \,,
    \qquad
    \delta\omega^{ij} = \tfrac12\bigl(\partial_j h_{ik}-\partial_i h_{jk}\bigr)\,dx^k
    + O(h^2) \,,
    \label{eq:delta_omega}
\end{equation}
where a prime denotes $\partial_\eta$; both are first order in $h$ and
carry a single derivative. (The general covariant form is
$\delta\omega^{ab}=\tfrac12(e^{a\mu}e^{b\nu}-e^{b\mu}e^{a\nu})
(\partial_\mu h_{\nu\rho}-\partial_\nu h_{\mu\rho})\vartheta^\rho$.)

The Clifford reduction is governed by the single identity
\begin{equation}
    \gamma^{ab}\gamma^c
    = \gamma^{abc} + \eta^{bc}\gamma^a - \eta^{ac}\gamma^b \,,
    \label{eq:clifford_split}
\end{equation}
which splits the product of the connection generator $\gamma^{ab}$
(carried by $\delta\omega=\tfrac14\delta\omega_{ab}\gamma^{ab}$) and the
frame generator $\gamma^c$ (carried by $\Psi=\vartheta^c\gamma_c\psi$)
into a \emph{vector} part $\eta^{bc}\gamma^a-\eta^{ac}\gamma^b$ and a
totally antisymmetric \emph{triple} $\gamma^{abc}$; no scalar,
pseudoscalar, or axial channel appears.

\paragraph{Cubic vertex.}
In the linear term
$2(\delta\omega\,\Psi\,\gamma_5\,\mathring D\Psi
+\mathring D\Psi\,\gamma_5\,\delta\omega\,\Psi)$ the connection enters
once. By~\eqref{eq:clifford_split} its $\gamma^{ab}$ contracts with the
frame $\gamma_c$ of the neighbouring $\Psi$; the antisymmetric triple
$\gamma^{abc}$ is annihilated against the symmetric $h_{ij}$, leaving
only the vector current. With $\delta\omega^{ij}\propto\partial h$
from~\eqref{eq:delta_omega} this collapses to
\begin{equation}
    \mathcal{L}_1 = -\frac{i}{2a^4}\,h_{ij}\,
    \bar\Psi_D\,\gamma^i\overset{\leftrightarrow}{\partial}_j\Psi_D \,,
    \label{eq:L1}
\end{equation}
the minimal graviton--fermion current, precisely the cubic vertex
$V_{h\psi\psi}$ of Ref.~\cite{MaleknejadKopp2026} [their Eq.~(4)]. Here
$\Psi_D$ denotes the Dirac spinor content of the QSL 1-form---the field
$\psi$ of $\Psi=\vartheta^a\gamma_a\psi$ [eq.~\eqref{eq:Psi_def}],
canonically normalized on the FRW background (the subscript $D$
distinguishes it from the Clifford-valued 1-form $\Psi$).

\paragraph{Quartic vertex.}
In the quadratic term $2\,\delta\omega\,\Psi\,\gamma_5\,\delta\omega\,\Psi$
the connection enters twice; applying~\eqref{eq:clifford_split} to each
factor, the two vector pieces recombine while the surviving
genuinely-new structure is the totally antisymmetric triple
$\gamma^{abc}$. With both $\delta\omega\propto\partial h$ this yields
\begin{equation}
    \mathcal{L}_2 = -\frac{i}{16a^3}\,
    e^\mu_\alpha\,h_{ij}\partial_\mu h_{ik}\,
    \bar\Psi_D\,\Gamma^{\alpha jk}\Psi_D \,,
    \label{eq:L2}
\end{equation}
reproducing the quartic vertex $V_{hh\psi\psi}$ [their Eq.~(5)], where
$\Gamma^{\alpha jk}\equiv\gamma^{\alpha jk}$ is the totally
antisymmetrized product of three gamma matrices.

The overall factor $i$ and the normalization are again those fixed by
the reality convention of $2D\Psi\gamma_5 D\Psi$
(section~\ref{sec:mass}); matching it to the coefficients
of~\eqref{eq:L1}--\eqref{eq:L2} shows the QSL reproduces the
minimal-coupling vertices. The advantage of the QSL derivation is that
equations~\eqref{eq:L1}--\eqref{eq:L2} follow from the single
algebraic split~\eqref{eq:clifford_split} \emph{without} selecting the
transverse-traceless gauge at the outset; gauge specialization can be
deferred to the final evaluation.

\section{Einstein--Cartan extension of the QSL}
\label{sec:EinsteinCartan}

The QSL papers~\cite{NesterTung1995} note that the formalism can be
generalized to Einstein--Cartan theory by allowing nonvanishing torsion.
In this section we carry out this generalization explicitly.

\subsection{QSL with torsion}

When torsion is allowed, $D\vartheta^a = T^a \neq 0$, and the
covariant derivative of $\Psi = \vartheta^a\gamma_a\psi$ becomes
\begin{equation}
    D\Psi = D(\vartheta^a\gamma_a\psi)
    = T^a\gamma_a\psi + \vartheta^a\gamma_a D\psi \,.
    \label{eq:DPsi_torsion}
\end{equation}
Define the ``torsion-free part'' $\mathring{D}\Psi := \vartheta^a\gamma_a\mathring{D}\psi$
(using the Levi-Civita connection $\mathring\omega$), and write the
full connection as
\begin{equation}
    \omega^{ab} = \mathring\omega^{ab} + K^{ab} \,,
    \label{eq:connection_split}
\end{equation}
where $K^{ab}$ is the contorsion 1-form.

The QSL Lagrangian becomes
\begin{equation}
    \mathcal{L}_{\rm QSL}
    = 2\,\mathring{D}\Psi\,\gamma_5\,\mathring{D}\Psi
    + 2\,\mathcal{K}\Psi\,\gamma_5\,\mathring{D}\Psi
    + 2\,\mathring{D}\Psi\,\gamma_5\,\mathcal{K}\Psi
    + 2\,\mathcal{K}\Psi\,\gamma_5\,\mathcal{K}\Psi \,,
    \label{eq:QSL_torsion}
\end{equation}
where $\mathcal{K} = \frac{1}{4}K_{ab}\gamma^{ab}$ is the
Clifford-algebra-valued contorsion.

\subsection{Field equations allowing torsion}

We now vary the action without imposing $T^a=0$. Because the
torsion-free derivative $\mathring D$ is built from the Levi-Civita
connection $\mathring\omega$, the contorsion $K^{ab}$ (equivalently
$\mathcal K=\tfrac14 K_{ab}\gamma^{ab}$) is an independent algebraic
variable, and $\delta_K(\mathring D\Psi)=0$,
$\delta_K(\mathcal K\Psi)=\delta\mathcal K\,\Psi$. Varying
\eqref{eq:QSL_torsion} with respect to $\mathcal K$ and using the
symmetry of the $\gamma_5$ bilinear,
\begin{equation}
    \frac{\delta\mathcal{L}_{\rm QSL}}{\delta\mathcal K}
    = \underbrace{2\bigl(\Psi\,\gamma_5\,\mathring D\Psi
      + \mathring D\Psi\,\gamma_5\,\Psi\bigr)}_{\text{from the linear (cross) terms}}
    + \underbrace{4\,(\mathcal K\Psi)\,\gamma_5\,\Psi}_{\text{from the }\mathcal K^2\text{ term}}
    = 0 \,.
    \label{eq:K_variation}
\end{equation}
This is the ``complete-the-square'' structure of
\eqref{eq:QSL_torsion}: the cross terms supply a source linear in the
spinor velocity $\mathring D\Psi$, while the $\mathcal K^2$ term supplies
the algebraic inertia. Equation~\eqref{eq:K_variation} is therefore
purely algebraic in $\mathcal K$ and is solved for the contorsion in
terms of the spinor and its torsion-free derivative. Re-expressed
through $K^{ab}\leftrightarrow T^a$, this is the torsion field equation
\begin{equation}
    \frac{\delta S}{\delta\omega^{ab}} = 0
    \quad\Longrightarrow\quad
    T^a = f(\psi, D\psi, \vartheta) \,.
    \label{eq:torsion_field_eq}
\end{equation}
The variation of $\Psi$ retains its earlier form $D^2\Psi=0$, but $D$
now carries the contorsion determined by~\eqref{eq:K_variation}, so this
is the Einstein--Cartan equation rather than the vacuum one.

In the standard Einstein--Cartan theory with Dirac matter, the torsion
is algebraically sourced by the spin density of the matter field.
Here the ``matter'' is the spinor factor $\psi$ of the gravitational
variable $\Psi$, and the torsion reflects the dynamics of the spinor
condensate.

\subsection{Cosmological spinor condensate and trace torsion}
\label{sec:EC:cosmological}

Consider a cosmological setting where the spinor field has the form
\begin{equation}
    \psi(t) = \chi(t)\,\hat\psi \,,
    \label{eq:spinor_condensate}
\end{equation}
with $\hat\psi$ a constant normalized spinor ($\bar{\hat\psi}\hat\psi=1$,
$\bar{\hat\psi}\gamma_5\hat\psi=0$) and $\chi(t)$ a real-valued
amplitude that evolves with cosmic time. (Throughout, $\chi(t)$ is this
scalar condensate amplitude; it is unrelated to the two-component spinor
1-form $\chi^A$ of section~\ref{sec:QSL:metric}, which appears only there.)

For the torsion-free QSL, the normalization $\bar\psi\psi = 1$ is
imposed as a constraint. In the Einstein--Cartan extension, we relax
this: $\chi(t)$ is a dynamical degree of freedom, and the normalization
condition is replaced by the torsion field equation.

The covariant derivative of the condensate~\eqref{eq:spinor_condensate} is
\begin{equation}
    D\psi = d\chi\,\hat\psi + \chi\,\omega\hat\psi
    = \chi\Bigl(\frac{d\chi}{\chi} + \omega\Bigr)\hat\psi
    = \chi\Bigl(d\ln\chi + \omega\Bigr)\hat\psi \,.
    \label{eq:Dpsi_condensate}
\end{equation}
In a spatially homogeneous background, $d\ln\chi = (\dot\chi/\chi)\,dt$
is purely timelike. This breaks the conformal equivalence between
$\chi(t)\hat\psi$ and a constant spinor, because the $d\ln\chi$ term
cannot be absorbed into a conformal rescaling of the tetrad alone.

\paragraph*{Why the source is the trace, not the axial, current.}
A general torsion 2-form decomposes into three Lorentz-irreducible
pieces~\cite{Hehl1976}---a trace (vector) part $V_a$, an axial (totally
antisymmetric) part $A_a$, and a traceless tensor part,
\begin{equation}
    T^a = \tfrac13\,V\wedge\vartheta^a
    + \tfrac16\,\epsilon^a{}_{bcd}\,A^b\,\vartheta^c\wedge\vartheta^d
    + {}^{(t)}T^a \,,
    \label{eq:torsion_irreducible}
\end{equation}
and the algebraic field equation~\eqref{eq:K_variation} maps each piece
to the corresponding irreducible part of the spinor source current. That
source is the cross-term 2-form
$J = \Psi\,\gamma_5\,\mathring D\Psi + \mathring D\Psi\,\gamma_5\,\Psi$.
Because the Levi-Civita derivative is torsion-free,
$\mathring D\vartheta^c = 0$, so $\mathring D\Psi=\vartheta^c\gamma_c\,
\mathring D\psi$, and with~\eqref{eq:Dpsi_condensate} (now read with
$\mathring\omega$),
\begin{equation}
    \mathring D\psi
    = \chi\bigl(\,\underbrace{d\ln\chi}_{\text{scaling, vector}}
    + \underbrace{\mathring\omega}_{\text{spin}}\,\bigr)\hat\psi \,,
    \label{eq:Dpsi_split}
\end{equation}
the source splits accordingly into two parts,
\begin{equation}
    J = J_{\rm scale} + J_{\rm spin},\qquad
    J_{\rm scale}\propto d\ln\chi \;\;(\text{vector}),\qquad
    J_{\rm spin}\propto n^a\equiv\bar{\hat\psi}\gamma_5\gamma^a\hat\psi
    \;\;(\text{axial}).
    \label{eq:source_split}
\end{equation}
This is the structural difference between the QSL and ordinary
Einstein--Cartan--Dirac theory. For a \emph{minimally coupled} Dirac
field the only spin current is the totally antisymmetric axial vector
$n^a$, which sources axial torsion alone~\cite{Hehl1976}. In the QSL the
gravitational variable $\Psi=\vartheta^a\gamma_a\psi$ carries the frame,
so the evolution of the condensate amplitude $\chi(t)$ contributes the
additional \emph{scaling} (dilatation) current $J_{\rm scale}$, built from
the timelike 1-form $d\ln\chi$ and absent for structureless matter; it
sources the \emph{vector} (trace) part of the torsion.

The cosmological symmetry then removes the axial part entirely. The axial
current is the constant 4-vector $n^a=\bar{\hat\psi}\gamma_5\gamma^a
\hat\psi$: its spatial components $n^i$ select a preferred direction and
so vanish by FRW isotropy, while its temporal component---the chirality
density---vanishes for the non-chiral condensate fixed
by~\eqref{eq:spinor_condensate} ($\bar{\hat\psi}\gamma_5\hat\psi=0$, equal
left/right content) on the parity-even background. Hence $J_{\rm spin}=0$
and only $J_{\rm scale}\propto d\ln\chi=(\dot\chi/\chi)\,dt$ survives.
The algebraic field equation~\eqref{eq:K_variation} equates each
Lorentz-irreducible part of the torsion to the corresponding part of this
source: with the vector piece $J_{\rm scale}$ the only nonzero channel, it
fixes the trace vector $V_a\propto(\dot\chi/\chi)\,\delta^0_a$ (the source
$d\ln\chi$ being purely timelike in the homogeneous background) and sets the
axial $A_a$ and traceless-tensor ${}^{(t)}T^a$ to zero. Through the
decomposition~\eqref{eq:torsion_irreducible} the torsion is therefore
purely the timelike trace vector,
\begin{equation}
    T^a = \tfrac13\,K\wedge\vartheta^a \,,\qquad A^a=0,\quad {}^{(t)}T^a=0,
    \label{eq:trace_torsion}
\end{equation}
with
\begin{equation}
    K_a = K\,\delta^0_a \,,\qquad K \propto \frac{\dot\chi}{\chi} \,.
    \label{eq:K_cosmological}
\end{equation}
The proportionality constant is an $O(1)$ number fixed by the torsion
field equation~\eqref{eq:K_variation}; we absorb it into $c_\chi$ in the
phenomenological estimates of section~\ref{sec:DM}. The vanishing of the
axial channel here is the geometric counterpart of the absence of a
pseudoscalar mass found in section~\ref{sec:mass}: both follow from the
same fact that the condensate sources a $V_a$, not an $A_a$.

\subsection{Contorsion for vectorial trace torsion}

For purely vectorial trace torsion~\eqref{eq:trace_torsion}, the
contorsion 1-form is~\cite{Hehl1976}
\begin{equation}
    K^{ab} = \frac{1}{3}\bigl(K^a\vartheta^b - K^b\vartheta^a\bigr) \,,
    \label{eq:contorsion}
\end{equation}
and the Clifford-algebra-valued contorsion is
\begin{equation}
    \mathcal{K} = \frac{1}{4}K_{ab}\gamma^{ab}
    = \frac{1}{6}K_a\,\vartheta_b\,\gamma^{ab}
    = \frac{1}{6}K_a(\gamma^a\vartheta_b\gamma^b - \vartheta^a) \,,
    \label{eq:K_clifford}
\end{equation}
where we used $\gamma^{ab}\vartheta_b = \gamma^a\vartheta_b\gamma^b
- \vartheta^a$, which follows from $\gamma^a\gamma^b=\gamma^{ab}+\eta^{ab}$.

\section{Spin content of the spinor 1-form}
\label{sec:spin32}

The spinor 1-form $\Psi_\mu = e^a_\mu\gamma_a\psi$ carries both a
spinor index (from $\psi$) and a spacetime vector index (from $e^a_\mu$).
We now analyze its spin content under the Lorentz group.

\subsection{Lorentz decomposition}

A Dirac-spinor-valued vector (or 1-form) transforms under the Lorentz
group as
\begin{equation}
    \bigl[(\tfrac{1}{2},0)\oplus(0,\tfrac{1}{2})\bigr]
    \otimes (\tfrac{1}{2},\tfrac{1}{2})
    = \underbrace{(\tfrac{1}{2},0)\oplus(0,\tfrac{1}{2})}_{\text{spin-}1/2}
    \oplus\underbrace{(1,\tfrac{1}{2})\oplus(\tfrac{1}{2},1)}_{\text{spin-}3/2} \,.
    \label{eq:Lorentz_decomposition}
\end{equation}
The spin-$1/2$ component is extracted by the gamma-trace:
\begin{equation}
    \psi^{(1/2)} \propto \gamma^\mu\Psi_\mu
    = \gamma^\mu e^a_\mu\gamma_a\psi = 4\psi \,,
    \label{eq:gamma_trace}
\end{equation}
and the spin-$3/2$ component is the gamma-traceless part:
\begin{equation}
    \Psi^{(3/2)}_\mu := \Psi_\mu - \frac{1}{4}\gamma_\mu\gamma^\nu\Psi_\nu \,.
    \label{eq:spin32_projection}
\end{equation}

A crucial point follows immediately. For the quadratic-spinor
1-form built from a single Dirac field, $\Psi_\mu=\gamma_\mu\psi$, the
gamma-traceless part vanishes identically,
\begin{equation}
    \Psi^{(3/2)}_\mu
    = \gamma_\mu\psi - \tfrac14\gamma_\mu\bigl(\gamma^\nu\gamma_\nu\bigr)\psi
    = \gamma_\mu\psi - \tfrac14\gamma_\mu(4\psi) = 0 \,,
    \label{eq:spin32_vanishes}
\end{equation}
where $\gamma^\nu\gamma_\nu = 4$ in four dimensions. This is not a
dimension-dependent accident: $\gamma_\mu\psi$ lies by construction in
the image of the gamma-trace map, so its gamma-traceless projection
vanishes in any dimension (the projector normalization $1/d$ cancels the
trace $\gamma^\nu\gamma_\nu=d$). Hence the QSL composite is purely the
spin-$1/2$ component: its sixteen apparent components are not independent
but are fixed by the four of $\psi$. The decomposition~\eqref{eq:Lorentz_decomposition} is the
representation content of a \emph{general} Dirac-vector $\Psi_\mu$; the
spin-$3/2$ sector is populated only when $\Psi_\mu$ is promoted to an
independent spinor-valued 1-form, with components beyond $\gamma_\mu\psi$.
\emph{In this paper we work with the composite field, so the propagating
excitation is the spin-$1/2$ Dirac field $\psi$.} The genuinely
independent spin-$3/2$ dynamics---a distinct theory with its own
constraint structure---is left to future work; section~\ref{sec:spin32ext} collects what the
kinematics and our later results suggest about that extension.

\section{Mass generation from trace torsion}
\label{sec:mass}

We now derive the effective mass that trace torsion confers on the
propagating spin-$1/2$ field $\psi$. By~\eqref{eq:spin32_vanishes} this
is the only dynamical degree of freedom of the composite QSL, so the
torsional term below is genuinely a Dirac mass for $\psi$, not a mass for
an independent spin-$3/2$ field.

\subsection{The $K^2$ term in the QSL}

From the torsion-expanded QSL~\eqref{eq:QSL_torsion}, the purely
torsional (mass-like) term is
\begin{equation}
    \mathcal{L}_{K^2}
    = 2\,\mathcal{K}\Psi\,\gamma_5\,\mathcal{K}\Psi \,.
    \label{eq:LK2}
\end{equation}
This term is algebraic in $\Psi$ (no derivatives) and quadratic in
the torsion $K$. It therefore has the structure of a mass term.

Substituting the contorsion~\eqref{eq:contorsion} for vectorial trace
torsion:
\begin{equation}
    \mathcal{K}\Psi
    = \frac{1}{4}K_{ab}\gamma^{ab}\,\vartheta^c\gamma_c\,\psi
    = \frac{1}{6}\bigl(K_a\vartheta_b - K_b\vartheta_a\bigr)
    \gamma^{ab}\,\vartheta^c\gamma_c\,\psi \,.
    \label{eq:KPsi}
\end{equation}

To evaluate~\eqref{eq:LK2} we write $\mathcal{K}\Psi$ as a
spinor-valued 2-form $\Xi = \mathcal{K}\wedge\Psi$, with frame
components
\begin{equation}
    \Xi_{ec}
    = \frac{1}{6}\,K_a\bigl(\gamma^{a}{}_{e}\gamma_c
      - \gamma^{a}{}_{c}\gamma_e\bigr)\psi \,,
    \qquad \gamma^a{}_e \equiv \gamma^{ab}\eta_{be} \,,
    \label{eq:Xi_components}
\end{equation}
so that the mass-like 4-form reduces to the spinor bilinear
\begin{equation}
    \mathcal{L}_{K^2}
    = \tfrac12\,\epsilon^{pqmc}\,
      \overline{\Xi_{pq}}\,\gamma_5\,\Xi_{mc}
    = \bar\psi\,\mathcal{S}\,\psi \,,
    \qquad
    \mathcal{S}
    = \tfrac12\,\epsilon^{pqmc}\,
      \overline{\Xi_{pq}}\,\gamma_5\,\Xi_{mc} \,.
    \label{eq:S_sandwich}
\end{equation}
The Clifford contractions are finite and unambiguous. Writing
$\Xi_{ec}=M_{ec}\psi$ with $M_{ec}=\tfrac16 K_a(\gamma^a{}_e\gamma_c
-\gamma^a{}_c\gamma_e)$, so that
$\mathcal{S}=\tfrac12\epsilon^{pqmc}\,\overline{M_{pq}}\,\gamma_5 M_{mc}$
($\overline{M}=\gamma^0M^\dagger\gamma^0$), and projecting onto the
complete covariant basis
$\{\mathbf{1},\gamma_5,\gamma_a,\gamma_5\gamma_a,\gamma_{ab}\}$ through
$\tfrac14\mathrm{tr}\,\mathcal{S}$,
$\tfrac14\mathrm{tr}(\gamma_5\mathcal{S})$, and so on, a direct
evaluation in a Dirac basis gives
\begin{equation}
    \mathrm{tr}\,\mathcal{S}\propto K^aK_a,\qquad
    \mathrm{tr}(\gamma_5\mathcal{S})=\mathrm{tr}(\gamma_a\mathcal{S})
    =\mathrm{tr}(\gamma_5\gamma_a\mathcal{S})
    =\mathrm{tr}(\gamma_{ab}\mathcal{S})=0 .
    \label{eq:S_traces}
\end{equation}
The pseudoscalar, vector, axial, and tensor channels all vanish
identically, and only the scalar channel survives,
\begin{equation}
    \mathcal{S} = \tfrac23\,K^aK_a\,\mathbf{1}
    \quad\Longrightarrow\quad
    \boxed{\ \mathcal{L}_{K^2}
    = \alpha\,K^aK_a\,\bar\Psi_\mu\Psi^\mu \,,\quad \alpha = \tfrac16\ }\,,
    \label{eq:LK2_expanded}
\end{equation}
where $\alpha$ is the coefficient of the scalar structure written in the
vector-spinor normalization, and we used
$\bar\Psi_\mu\Psi^\mu = \bar\psi\gamma_\mu\gamma^\mu\psi = 4\,\bar\psi\psi$,
so that the coefficient of $\bar\psi\psi$ is $4\alpha=\tfrac23$ and
$\tfrac14\mathrm{tr}\,\mathcal{S}=\tfrac23K^aK_a$.
The vanishing of the pseudoscalar trace [eq.~\eqref{eq:S_traces}] means
there is no $\gamma_5$ mass mixing, and the vanishing of the remaining
traces means there are \emph{no} cross terms: the trace-torsion mass term
is a pure Dirac scalar. The overall factor $i$ generated by the
$\gamma_5$--$\epsilon$ pairing
($\gamma_5 \propto i\,\epsilon_{abcd}\gamma^{abcd}$) is the universal
reality convention of the QSL---it multiplies the kinetic term
identically and is absorbed in the normalization that renders
$2D\Psi\gamma_5 D\Psi$ the real Einstein--Hilbert
density~\eqref{eq:spinor_curvature_identity}; it therefore cancels in
the physical mass and does not affect $\alpha$. The positivity
$\alpha>0$ follows because $\mathcal{L}_{K^2}$ is the same
$2\,\overline{X}\gamma_5 X$ quadratic form, evaluated on the
timelike-sourced 2-form $X=\mathcal{K}\Psi$, whose definiteness underlies
the Nester--Tung positive-energy property of the QSL.

Using $\bar\Psi_\mu\Psi^\mu=4\bar\psi\psi$, the result~\eqref{eq:LK2_expanded}
is explicitly a Dirac mass term for the spin-$1/2$ field,
\begin{equation}
    \mathcal{L}_{K^2}
    = 4\alpha\,K^aK_a\,\bar\psi\psi
    = -\tfrac{2}{3}K^2\,\bar\psi\psi \,,
    \label{eq:dirac_mass}
\end{equation}
for the cosmological timelike torsion $K^aK_a=-K^2$. That the result is a
scalar ($\bar\psi\psi$) rather than the axial four-fermion contact term
$(\bar\psi\gamma_5\gamma_a\psi)^2$ familiar from Einstein--Cartan theory
with Dirac matter~\cite{Hehl1976} is a consequence of the source: here
the torsion is the \emph{vectorial trace} torsion carried by the
homogeneous condensate, a c-number background $K_a\propto\dot\chi/\chi$,
not the axial spin density of a propagating Dirac field. The mass is thus
an effective mass acquired by fluctuations of $\psi$ in the torsionful
condensate background, and the vanishing pseudoscalar channel guarantees
it is a pure scalar mass with no axion-like component.

\subsection{Effective mass matrix}

In the cosmological setting of section~\ref{sec:EC:cosmological},
$K_a = K\delta^0_a$ with $K \propto \dot\chi/\chi$, so
$K^aK_a = -K^2 < 0$ (timelike vector, mostly-plus convention).
With $\alpha=\tfrac16$ from~\eqref{eq:LK2_expanded}, the effective
mass-squared is therefore
\begin{equation}
    M^2_{\rm eff} = -\alpha\,K^aK_a
    = \alpha\,K^2
    = \frac{1}{6}\left(\frac{\dot\chi}{\chi}\right)^2 > 0 \,,
    \label{eq:mass_squared}
\end{equation}
manifestly positive: the geometric-drag mechanism produces a genuine
(non-tachyonic) mass. Here $M^2_{\rm eff}$ is read off in the same
vector-spinor normalization as the QSL kinetic term
$2\,\mathring{D}\Psi\gamma_5\mathring{D}\Psi$, i.e. relative to the
bilinear $\bar\Psi_\mu\Psi^\mu$; the common factor $4=\gamma_\mu\gamma^\mu$
that converts between $\bar\Psi_\mu\Psi^\mu$ and $\bar\psi\psi$ cancels in
the mass-to-kinetic ratio, leaving the coefficient $\alpha=\tfrac16$ exact.
Equivalently, the second-order field equation $\mathring{D}^2\Psi$ acquires
the algebraic shift $\mathring{D}^2\Psi\to\mathring{D}^2\Psi+M^2_{\rm eff}\Psi$
[cf.~eq.~\eqref{eq:dispersion}], which fixes the same $M^2_{\rm eff}=\alpha K^2$.

\subsection{Physical interpretation: geometric drag}

The mass generation mechanism has a transparent physical interpretation.
As the spinor condensate amplitude $\chi(t)$ evolves in the expanding
universe, it generates trace torsion $K \propto \dot\chi/\chi$. This
torsion acts as a ``geometric drag'' on the spin-$1/2$ field $\psi$,
coupling to it through the shared spin connection. The result is an
effective mass that is cosmologically generated---it vanishes in a
static universe ($\dot\chi = 0$) and grows as the condensate evolves.

This mechanism eliminates the need for a separate Higgs sector to
generate the DM mass required by Ref.~\cite{MaleknejadKopp2026}.
The mass arises from the same geometric dynamics that also produces
the stochastic perturbations responsible for the freeze-in mechanism.

\subsection{Decoupling from transverse-traceless modes}
\label{sec:torsion_TT}

It remains to ask how the trace torsion relates to the
transverse-traceless (TT) metric perturbations $h_{ij}^{\rm TT}$ that
carry the gravitational waves of the production mechanism. Because the
torsion enters the QSL quadratically (section~\ref{sec:mass}), any
coupling to a single graviton is mediated, at lowest order, by the
spin-$2$ part of the bilinear $K_aK_b$. We decompose this bilinear under
the $SO(3)$ rotation group of the spatial slices.

For the cosmological condensate the trace-torsion vector is purely
timelike, $K_a=K\delta^0_a$, so its spatial components vanish,
$K_i=0$, and the spatial source tensor is identically zero,
\begin{equation}
    K_iK_j = 0 \,,
    \qquad
    \Lambda_{ij}{}^{kl}\,K_kK_l = 0 \,,
    \label{eq:no_TT_source}
\end{equation}
where $\Lambda_{ij}{}^{kl}=P_i{}^kP_j{}^l-\tfrac12 P_{ij}P^{kl}$,
$P_{ij}=\delta_{ij}-\hat k_i\hat k_j$, is the TT projector for a mode of
direction $\hat k$. The timelike bilinear contributes only the
$SO(3)$ scalar $K_0K_0=K^2$ and the isotropic spatial piece
$\propto\delta_{ij}$; since $\Lambda_{ij}{}^{kl}\delta_{kl}=0$, neither
has a TT component. \emph{Homogeneous trace torsion therefore does not
source transverse-traceless gravitational waves at second order}: the
scalar (mass) sector of section~\ref{sec:mass} and the tensor (GW)
sector are orthogonal representations and do not mix. This is the
representation-theoretic counterpart of the causal result of
section~\ref{sec:characteristic}---there is no torsion-induced
birefringence or tilting of the TT light cone, precisely because the
torsion has no spin-$2$ projection to feed into it.

Two channels could in principle open the coupling, and the same
projector shows which survive. A spatial \emph{gradient} of a torsion
scalar, entering as $\partial_i\partial_j(K^2)$, is purely longitudinal,
$\Lambda_{ij}{}^{kl}\hat k_k\hat k_l=0$, and still yields no TT source.
Only genuinely \emph{anisotropic} torsion with transverse spatial
components $K_i$ produces a nonzero TT part: for example a torsion
vector $K_i=K\hat x_i$ probed by a wave along $\hat z$ gives
\begin{equation}
    \Lambda_{ij}{}^{kl}K_kK_l
    = \tfrac12 K^2\,(\hat x_i\hat x_j - \hat y_i\hat y_j) \,,
    \label{eq:TT_source_aniso}
\end{equation}
a pure ``$+$'' polarization. Observable GW signatures of QSL torsion
thus require departures from spatial isotropy; the homogeneous
condensate that generates the dark-matter mass is, by construction,
gravitationally silent in the tensor channel.

\section{Dark matter phenomenology}
\label{sec:DM}

\subsection{Production mechanism}

In the QSL framework, the gravitational-wave induced freeze-in
of Ref.~\cite{MaleknejadKopp2026} is reinterpreted as follows.
Both the GW perturbation and the produced fermions are excitations
of the same spinor 1-form $\Psi$. The GW corresponds to a perturbation
$\delta\Psi$ of the gravitational sector, while the produced fermion
is an excitation of the matter (spin-$1/2$) sector. Their coupling
is mediated by the shared spin connection.

In the super-$SL(2,\mathbb{C})$ language of section~\ref{sec:QSL:super},
the production mechanism is a dynamical realization of the anticommutation
relation $\{Q_A,Q_B\} = 2M_{AB}$: gravitational excitations (associated
with $M_{AB}$) source fermionic modes (associated with $Q_A$).

\subsection{Mass scale}

The effective mass~\eqref{eq:mass_squared} is set by $K \propto
\dot\chi/\chi$. In the early universe the condensate evolves on a
Hubble time, $\dot\chi/\chi = c_\chi H$ with $c_\chi=O(1)$, so
with $\alpha=\tfrac16$ from~\eqref{eq:LK2_expanded} the mass is locked
to the expansion rate at the production epoch $H_*$,
\begin{equation}
    M_{\rm eff} = \sqrt{\alpha}\,\Bigl|\frac{\dot\chi}{\chi}\Bigr|
    = \frac{c_\chi}{\sqrt6}\,H_* \;\simeq\; 0.4\,c_\chi\,H_* \,.
    \label{eq:mass_scale}
\end{equation}
Here $\alpha=\tfrac16$ is the exact coefficient of the Clifford
reduction~\eqref{eq:LK2_expanded} at fixed normalization of the
contorsion, whereas $c_\chi\equiv\dot\chi/(\chi H)$ is the undetermined
$O(1)$ ratio set by the condensate dynamics; the precise mass-to-Hubble
ratio is therefore $c_\chi/\sqrt6$, not $1/\sqrt6$, and fixing $c_\chi$
requires the explicit solution of~\eqref{eq:K_variation} in a given
cosmological model. This is the central qualitative feature of the QSL
mechanism: unlike a generic dark-matter model, the mass is \emph{not} a
free parameter but is tied to the Hubble scale at production. As the
universe expands and
the condensate approaches its asymptotic value, $\dot\chi/\chi\to0$ and
the mass freezes out at the value~\eqref{eq:mass_scale} evaluated at
$t_{\rm freeze}$.

\subsection{Relic density}

\begin{figure}[t]
\centering
\includegraphics[width=0.78\linewidth]{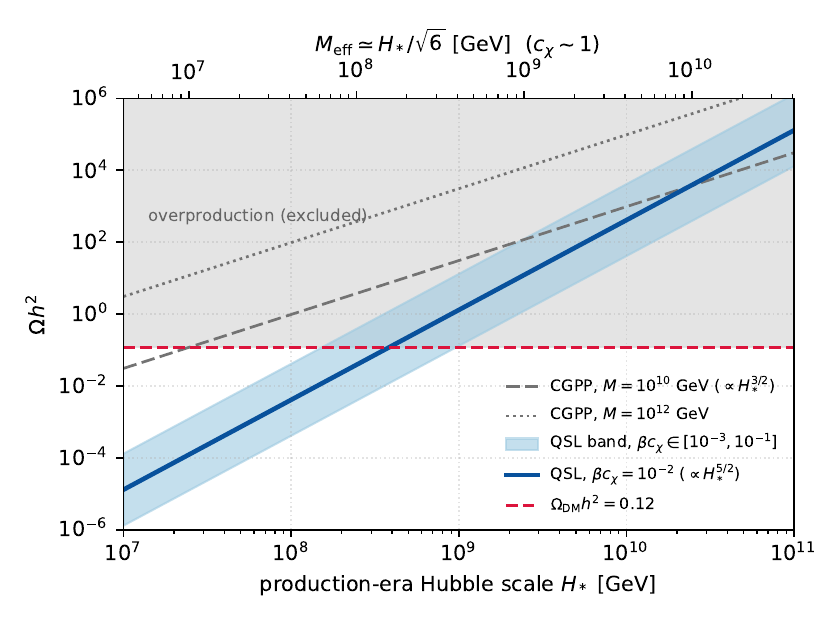}
\caption{Relic abundance of the gravitationally produced spin-$1/2$
fermion as a function of the production-era Hubble scale $H_*$, from the
mass-locking scaling~\eqref{eq:relic}, $\Omega h^2\propto H_*^{5/2}$
(blue). The band spans the $O(1)$ uncertainty
$\beta c_\chi\in[10^{-3},10^{-1}]$ in the yield and condensate
parameters; the solid line is $\beta c_\chi=10^{-2}$. The observed value
$\Omega_{\rm DM}h^2=0.12$ (dashed red) selects an intermediate production
scale, $H_*\sim10^{8}$--$10^{9}\,$GeV, and hence a geometrically fixed
mass $M_{\rm eff}\simeq H_*/\sqrt6$ (top axis, shown for $c_\chi\sim1$);
abundances in the shaded region overclose the universe. For contrast, the
grey lines show generic cosmological gravitational production with an
\emph{independent} mass, $\Omega h^2\propto H_*^{3/2}$ at fixed $M$
(shown for $M=10^{10}$ and $10^{12}\,$GeV): there the mass and $H_*$ are
separate inputs, a two-parameter family. Mass locking changes the slope
$\tfrac32\to\tfrac52$ and collapses that family onto the single QSL curve.
The figure is illustrative, since the prefactor still depends on $\beta$,
$c_\chi$, and the reheating history.}
\label{fig:relic}
\end{figure}

The mass-locking relation~\eqref{eq:mass_scale} turns the relic-abundance
estimate into a one-parameter prediction. The dark-matter candidate is
the spin-$1/2$ Dirac field $\psi$---the same fermionic content as the
Weyl fermion of Ref.~\cite{MaleknejadKopp2026}---so the gravitational
production rate is the well-controlled spin-$1/2$ case. For production at
the end of inflation the comoving number density is $n_*=\beta\,H_*^3$,
where the dimensionless fermionic yield $\beta$ depends on
$M_{\rm eff}/H_*$; since~\eqref{eq:mass_scale} places
$M_{\rm eff}\lesssim H_*$, one is in the unsuppressed regime
$\beta\sim10^{-3}$--$10^{-1}$~\cite{KolbLong2024}. With (near-)instant
reheating, $T_{\rm RH}\simeq(90/\pi^2 g_*)^{1/4}\sqrt{H_*M_{\rm Pl}}$,
the yield $Y=n_*/s$ and $\Omega h^2 = 2.74\times10^{8}\,(M_{\rm eff}/{\rm GeV})\,Y$
give
\begin{equation}
    \Omega_{\rm DM}h^2
    \;\simeq\; 1.3\times10^{12}\;\beta\,c_\chi
    \left(\frac{H_*}{10^{13}\,{\rm GeV}}\right)^{5/2} ,
    \label{eq:relic}
\end{equation}
where the power $\tfrac52$ rather than the usual $\tfrac32$ is the direct
imprint of mass locking ($M_{\rm eff}\propto H_*$, so
$\Omega\propto M_{\rm eff}H_*^{3/2}\propto H_*^{5/2}$). This monotonic
dependence on a single scale is the distinctive signature of the QSL:
matching the observed $\Omega_{\rm DM}h^2\simeq0.12$ fixes the production
scale and hence the mass. For $\beta c_\chi\sim10^{-2}$ this gives
\begin{equation}
    H_*\sim4\times10^{8}\,{\rm GeV} \,,
    \qquad
    M_{\rm eff}\sim1.6\times10^{8}\,{\rm GeV} \,,
    \label{eq:relic_number}
\end{equation}
an intermediate-scale superheavy candidate. We stress that
\eqref{eq:relic_number} is no more than indicative. The yield $\beta$ used
here is borrowed from \emph{generic} cosmological gravitational
production~\cite{KolbLong2024}; the actual value for the
gravitational-wave-induced channel of Ref.~\cite{MaleknejadKopp2026}
requires its full mode-function computation and may differ. Moreover,
because $\Omega h^2\propto H_*^{5/2}$, the order-of-magnitude span
$\beta c_\chi\sim10^{-3}$--$10^{-1}$ maps to only a factor
$\sim10^{4/5}\approx6$ in $H_*$---so the production scale is bounded but
hardly pinned, and the single pair of numbers in~\eqref{eq:relic_number}
should be read as the centre of a broad band, not a prediction. The
robust, model-independent outputs are the locking
relation~\eqref{eq:mass_scale} and the scaling~\eqref{eq:relic},
illustrated in Fig.~\ref{fig:relic}.

\subsection{Comparison with other gravitational dark-matter scenarios}

The QSL candidate---a spin-$1/2$ Dirac fermion with a geometrically fixed
mass---is most naturally compared with the parent Weyl-fermion scenario
of Ref.~\cite{MaleknejadKopp2026} and the broader programme of
cosmological gravitational particle production
(CGPP)~\cite{KolbLong2024}. Its distinguishing feature is mass locking:
in CGPP the mass $M$ and the production-era Hubble scale $H_*$ are
independent inputs scanned to fit $\Omega_{\rm DM}h^2$, whereas the QSL
fixes $M_{\rm eff}\simeq(c_\chi/\sqrt6)H_*$ [\eqref{eq:mass_scale}],
collapsing the two-parameter $(M,H_*)$ plane to the one-parameter
line~\eqref{eq:relic} (at fixed $c_\chi$). The mass is not postulated, as in the
Maleknejad--Kopp construction, but supplied by the trace torsion of the
same condensate. The candidate also sits within the broader
Einstein--Cartan dark-matter programme, in which torsion-induced
four-fermion interactions \emph{produce} singlet-fermion dark matter over a
wide mass range~\cite{ECportal}; the QSL plays a complementary role, with
the trace torsion fixing the produced fermion's mass and spin rather than
its production rate. Viewed through the super-$SL(2,\mathbb{C})$ structure of the
QSL, this spin-$1/2$ candidate is the Goldstino of the local supersymmetry broken
by the metric condensate $\langle\bar\Psi\Psi\rangle$, and so descends from the
oldest supersymmetric dark-matter idea---the light gravitino of
Pagels and Primack~\cite{PagelsPrimack1982}, whose helicity-$1/2$ (Goldstino)
component dominates---but as a \emph{composite} Goldstino of \emph{gravitational}
supersymmetry, with a geometric mass and no superpartner spectrum (developed in
the companion paper~\cite{TungNoGo}).

It is worth situating the present mechanism against the broader use of
torsion for dark matter, which has followed three distinct routes. In
quadratic Poincar\'e gauge gravity (``$R+T^2$''), quadratic torsion
invariants make the torsion \emph{propagate}, and a stable pseudoscalar
(spin-$0^-$) tordion can account for cold dark
matter~\cite{CruzDombrizDM}; such propagating-torsion theories are,
however, governed throughout by the ghost/tachyon problem, with stability
generically forcing the torsion to become non-dynamical and the theory to
collapse back to general relativity~\cite{PGTstability}. A second route is
the Einstein--Cartan portal already noted~\cite{ECportal}, with
non-dynamical torsion. A third, closest in phenomenology to the present
work, is the ``steady-state torsion'' line of Friedmann
cosmologies~\cite{KranasTorsion,PereiraTorsionDM}, in which a
\emph{vectorial} (trace) torsion $S_\alpha=-3\phi\,u_\alpha$ with
$\phi=-\alpha H\propto H$ modifies the expansion law to mimic an
\emph{effective} dark-matter density---the same trace-torsion-$\propto H$
scaling realised here by $K\propto\dot\chi/\chi$, but introduced as a
phenomenological ansatz and yielding no dark-matter particle. The QSL
mechanism is distinct from all three: the trace torsion is not posited but
\emph{derived} from the geometry-building spinor condensate; it is
algebraic (non-dynamical), so no propagating-torsion ghost arises; and it
does not produce an effective fluid but \emph{generates the mass of a
genuine composite spin-$\tfrac12$ particle}, the unique propagating QSL
excitation (section~\ref{sec:spin32}). QSL thus shares the vectorial
torsion-$\propto H$ phenomenology of the steady-state line while supplying
what that line lacks---a first-principles origin and an actual dark-matter
quantum---and avoids the ghost burden of the propagating-torsion ($R+T^2$)
route.

The future spin-$3/2$ extension (section~\ref{sec:spin32}) would instead
make contact with the ``raritron'' gravitational-production
analyses~\cite{KanetaKeMambriniOliveVerner2023} and the gravitino
literature~\cite{CheungElorHall2011}. There, massive spin-$3/2$ fields
generically suffer a catastrophic enhancement of the longitudinal
($s=\tfrac12$, Goldstino-like) mode, whose vanishing sound speed drives
overproduction and signals a low cutoff, and which the raritron
literature tames by a suitably \emph{time-dependent} mass. It is
suggestive that the QSL would supply exactly such a mass,
$M_{\rm eff}(t)\propto\dot\chi/\chi$; we make this precise
in section~\ref{sec:longitudinal} as a motivation for the extension,
while stressing that it bears on the independent-field theory, not on the
spin-$1/2$ candidate of the present paper.
\subsection{Gravitational-wave signatures}
\label{sec:gwsig}

The locking relation yields two quantitative, and distinctive,
gravitational-wave (GW) signatures. The first is a \emph{production peak}.
The production scale $H_*$ fixes \emph{both} the mass and the redshifted
horizon-scale frequency of the gravitational waves present at production,
\begin{equation}
  f_0=\frac{a_*}{a_0}\frac{H_*}{2\pi},\qquad
  \frac{a_*}{a_0}=\Bigl(\frac{g_{s0}}{g_{s*}}\Bigr)^{1/3}\frac{T_0}{T_*},\quad
  3M_{\rm Pl}^2H_*^2=\frac{\pi^2}{30}g_*T_*^4 ,
\end{equation}
so that, with $M_{\rm eff}\propto H_*$, the peak frequency is \emph{locked
one-to-one to the dark-matter mass}, $f_0\propto\sqrt{M_{\rm eff}}$. Across the
relic-viable window $H_*\sim10^{7}$--$10^{10}\,$GeV this gives
$f_0\simeq0.07$--$2\,$MHz (e.g.\ $f_0\simeq0.44\,$MHz at the fiducial
$M_{\rm eff}\simeq1.6\times10^{8}\,$GeV)---the ultra-high-frequency (UHF) band of
bulk acoustic-wave antennas and resonant cavities, above ground-based
interferometers and below GHz haloscopes. The mass--frequency locking is the
sharp signature: no astrophysical UHF source correlates its peak frequency with a
dark-matter mass.

\begin{figure}[t]
  \centering
  \includegraphics[width=0.86\linewidth]{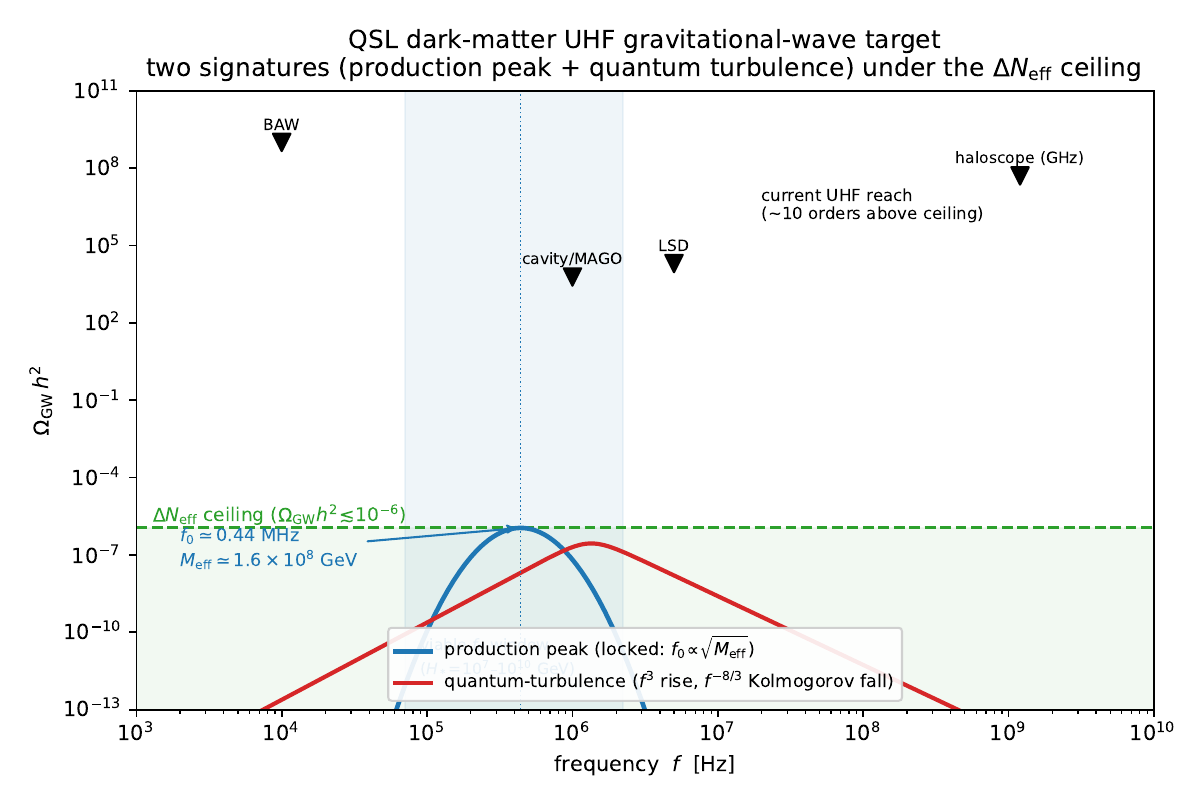}
  \caption{QSL dark-matter UHF gravitational-wave target. The production peak
  (blue) sits at $f_0\propto\sqrt{M_{\rm eff}}$ (fiducial $0.44\,$MHz; viable band
  shaded); the quantum-turbulence background (red) is broadband
  ($f^3$ rise, $f^{-8/3}$ Kolmogorov fall). Both lie under the
  $\Delta N_{\rm eff}$ ceiling (green). Current UHF sensitivities (markers) remain
  far above, so the prediction is a specific target for future UHF detectors.}
  \label{fig:gwtarget}
\end{figure}

A second, shape-distinct background arises because the dark sector is a superfluid
spinor condensate, whose relaxation proceeds through \emph{quantum turbulence}
(Sec.~\ref{sec:turbulence}). A turbulent vortex tangle radiates a broadband
stochastic spectrum---a causal $\Omega_{\rm GW}\propto f^3$ rise below the
integral scale and a Kolmogorov $\propto f^{-8/3}$ fall above it---peaked near a
few times $f_0$. The combined QSL prediction is therefore a two-component UHF
spectrum (Fig.~\ref{fig:gwtarget}): a narrow production peak \emph{at} $f_0$
riding a Kolmogorov turbulence shoulder, a composite shape that no single
astrophysical line or flat cosmological background reproduces.

Both components are relativistic relics and so are bounded by the radiation
budget, $\Omega_{\rm GW,0}h^2\lesssim5.6\times10^{-6}\,\Delta N_{\rm eff}\simeq
10^{-6}$. We stress the honest status: present UHF detectors lie several orders of
magnitude above this ceiling, so the signal is a target for the developing UHF-GW
programme rather than for current instruments---but it is a \emph{specific} one,
fixing which dark-matter mass each UHF band probes and predicting a definite
two-component spectral shape, so that measuring the dark-matter mass would fix the
frequency of its stochastic gravitational-wave counterpart, and conversely.

\section{Connection to scalar field cosmology}
\label{sec:cosmology}

\subsection{Halpern--Huang scalar field and asymptotic freedom}

Huang, Low and Tung~\cite{HuangLowTung2012a,HuangLowTung2012b} built a
cosmology on the Halpern--Huang asymptotically free scalar
potential---the potential is due to Halpern and Huang, while its
cosmological use and the accompanying trace-anomaly term were introduced
in these works. The key results are: (i) the Hubble parameter follows a
power law $H \sim t^{-p}$; (ii) the effective cosmological constant
decays as $t^{-2p}$, avoiding the fine-tuning problem; (iii) the
trace-anomaly term $\Lambda\,\partial V/\partial\Lambda = \beta(V)$,
introduced by Huang, Low and Tung, breaks scale invariance and drives
the expansion.

\subsection{Trace anomaly, trace torsion, and the dark-matter mass}

The Huang--Low--Tung trace anomaly is not merely analogous to the
conformal breaking of the QSL mechanism---it is its source, and it is
carried geometrically by the very trace torsion that generates the
dark-matter mass. Three statements make this precise.

\paragraph{The anomaly is the origin of the mass.}
The effective mass $M_{\rm eff}=\tfrac{1}{\sqrt6}|\dot\chi/\chi|$
[\eqref{eq:mass_scale}] is nonzero only because the condensate evolves,
i.e.\ only because scale invariance is broken. A scale-invariant
condensate sits at a fixed point with $\dot\chi/\chi=0$ and is massless.
In the Huang--Low--Tung cosmology the breaking is the trace anomaly
$T^\mu{}_\mu=\beta(V)=\Lambda\,\partial V/\partial\Lambda$, so the anomaly
is precisely what drives $\dot\chi/\chi\neq0$ and hence what endows the
dark matter with mass. The ``geometric drag'' of section~\ref{sec:mass}
and the trace anomaly are two descriptions of the same scale breaking.

\paragraph{Trace torsion as the dilatation connection.}
The vectorial (trace) part of the torsion is the geometric carrier of
local scale (dilatation) transformations: its nonvanishing is the
geometric face of a broken dilatation symmetry, just as the
non-conservation of the dilatation current
$\partial_\mu j^\mu_{\rm D}=T^\mu{}_\mu=\beta(V)$ is its field-theoretic
face. The QSL generates exactly a vectorial trace torsion
$K_a\propto\dot\chi/\chi$, giving the correspondence
\begin{equation}
    K_a \;\longleftrightarrow\; \text{dilatation connection}\,,
    \qquad
    \nabla^a K_a \;\longleftrightarrow\; T^\mu{}_\mu=\beta(V)\,.
    \label{eq:dilatation_correspondence}
\end{equation}
We read~\eqref{eq:dilatation_correspondence} as a \emph{current-level
correspondence}, not a strict geometric identity: it expresses the
parallel roles played by the Maleknejad--Kopp conformal breaking by
gravitational waves, the Huang--Low--Tung trace anomaly, and the QSL
trace torsion in supplying the same scale-symmetry breaking, and we
suggest these are three facets of one mechanism. The caveat is that the
canonical Weyl/dilatation vector properly resides in the non-metricity
and coincides with the trace torsion only up to a projective frame
redefinition; establishing~\eqref{eq:dilatation_correspondence} as a
geometric identity, rather than a matching of divergences, is left to
future work.

\paragraph{The anomaly would fix the prediction.}
The relic estimate of section~\ref{sec:DM} carried one undetermined
$O(1)$ ratio, $c_\chi=\dot\chi/(\chi H)$. The Huang--Low--Tung trace
anomaly $\beta(V)$ on the Halpern--Huang asymptotically free potential is
what sets the rolling rate $\dot\chi/\chi$, and would thereby fix $c_\chi$
and---through the
locking~\eqref{eq:mass_scale} and the scaling~\eqref{eq:relic}---the
mass and relic abundance. The freeze-out of the mass as
$\dot\chi/\chi\to0$ is the flow toward the infrared fixed point at which
the anomaly switches off and scale invariance is restored. The trace
anomaly thus governs both the magnitude and the time profile of
$M_{\rm eff}$; making the map $\beta(V)\to c_\chi$ quantitative within a
definite Halpern--Huang background is the natural route to a firm relic
prediction.

\subsection{Stochastic perturbations from quantum turbulence}
\label{sec:turbulence}

The Maleknejad--Kopp mechanism requires a second external input besides
the mass: a stochastic gravitational-wave background. The same
Huang--Low--Tung condensate supplies it. A complex scalar with spatially
varying phase behaves as a superfluid, and its
flow---$v_a\propto\partial_a\theta$ for phase $\theta$---develops quantum
turbulence~\cite{HuangLowTung2012b}, with a Kolmogorov inertial-range
cascade $E(k)\propto k^{-5/3}$ between the integral (coherence) scale and
the dissipation scale. The transverse-traceless part of the turbulent
anisotropic stress radiates gravitational waves, a conversion worked out
in detail for cosmological
turbulence~\cite{KosowskyMackKahniashvili2002,CapriniDurrer2006}: the
resulting background is peaked near the integral scale---of order the
horizon, $\ell\sim H_*^{-1}$, at the production epoch---with amplitude
set by the fraction of the energy density in turbulent motion, and
power-law tails inherited from the cascade.

This dovetails with the tensor-sector analysis of
section~\ref{sec:torsion_TT}. There we found that the homogeneous,
isotropic trace torsion is gravitationally silent in the
transverse-traceless channel: the spin-$2$ projection of $K_aK_b$
vanishes, so the mass-generating condensate sources no GWs by itself, and
a transverse-traceless signal requires an \emph{anisotropic} source. The
turbulent velocity field is exactly such a source---inhomogeneous and
anisotropic by construction. The two roles of the condensate are thus
complementary and non-competing: its homogeneous mode generates the
trace torsion and hence the dark-matter mass
(sections~\ref{sec:EinsteinCartan}--\ref{sec:mass}), while its turbulent
fluctuations generate the transverse-traceless background that drives the
freeze-in production. A single condensate therefore furnishes both inputs
that the Maleknejad--Kopp mechanism takes as given, and the trace anomaly
that breaks scale invariance is the common origin of both.

A quantitative prediction---the GW amplitude and spectral shape at the
characteristic high frequencies set by the production scale $H_*$, and
their correlation with the dark-matter mass through that shared scale
[$f_0\propto\sqrt{M_{\rm eff}}$, section~\ref{sec:DM}]---requires evolving
the Huang--Low--Tung turbulence spectrum through the conversion
of~\cite{KosowskyMackKahniashvili2002,CapriniDurrer2006}.
This is the most direct observational handle on the framework and the
natural next step.

\section{Toward an independent spin-3/2 field}
\label{sec:spin32ext}

Throughout, the propagating field has been the spin-$1/2$ Dirac fermion
$\psi$ of the composite QSL. It is natural to ask about the
\emph{independent} spin-$3/2$ extension, in which $\Psi_\mu$ is promoted to
a vector-spinor with components beyond $\gamma_\mu\psi$. We have since
settled this question in Paper~II~\cite{TungNoGo}, with a
\emph{negative} answer: the independent field acquires no propagating
massive spin-$3/2$ mode, because the QSL action depends on $\Psi$ only
through the metric $g=\Psi\otimes_S\Psi$ and the scalar $\bar\psi\psi$, so
its quadratic fluctuations factor through a massless graviton and scalar.
We summarize here the structural analyses that led to that result and
indicate where each enters the no-go, separating what is established from
what was conditional. The one outright result is the
principal-symbol/characteristic-determinant analysis
(section~\ref{sec:characteristic}): the linearized QSL is second order
with a metric-diagonal principal symbol, so its leading characteristic
cone is the metric cone---this holds for the full reducible multiplet,
does not rely on any Rarita--Schwinger input, and is the statement, in
Paper~II, that all propagation lies on the metric light cone. The
remaining analyses---the Rarita--Schwinger comparison (section~\ref{sec:RS}),
the helicity-$\tfrac12$ sound speed (section~\ref{sec:longitudinal}), and
the cosmological constraint reduction (section~\ref{sec:constraints})---were
\emph{conditional}: they import first-order Rarita--Schwinger structures
whose transfer to the second-order QSL is a working assumption, flagged at
each step. They are superseded by Paper~II, which needs no
such import; we retain them as motivation and as the constraint-sector
description of why the would-be spin-$3/2$ modes are non-dynamical.

\subsection{Comparison with the Rarita--Schwinger field}
\label{sec:RS}

The standard Rarita--Schwinger Lagrangian for a free massive spin-$3/2$
field $\Psi_\mu$ is~\cite{RaritaSchwinger1941}
\begin{equation}
    \mathcal{L}_{\rm RS}
    = -\epsilon^{\mu\nu\rho\sigma}\,
    \bar\Psi_\mu\gamma_5\gamma_\nu D_\rho\Psi_\sigma
    - m\,\bar\Psi_\mu\Sigma^{\mu\nu}\Psi_\nu \,,
    \label{eq:RS_Lagrangian}
\end{equation}
where $\Sigma^{\mu\nu} = g^{\mu\nu} - \gamma^\mu\gamma^\nu$.
The kinetic term is first-order in derivatives; the mass term is algebraic.

The QSL, written schematically as
$\mathcal{L}_{\rm QSL} = 2\,(D_\mu\Psi_\nu)\gamma_5(D^\mu\Psi^\nu)$
(suppressing form-degree structure), is second-order in derivatives,
whereas the Rarita--Schwinger kinetic term is first-order. The mass term
in~\eqref{eq:RS_Lagrangian} is proportional to $\bar\Psi_\mu\Psi^\mu$,
and trace torsion generates precisely this contraction in the QSL, as
shown in section~\ref{sec:mass}.

The relation must, however, be stated carefully in view
of~\eqref{eq:spin32_vanishes}. The composite QSL field carries no
independent spin-$3/2$ component, so~\eqref{eq:RS_Lagrangian} is
\emph{not} the dynamics realized here; the QSL realizes the spin-$1/2$
trace, and the $\bar\Psi_\mu\Psi^\mu$ contraction it generates reduces
to a Dirac mass for $\psi$ (section~\ref{sec:mass}). The kinematic
resemblance to a Rarita--Schwinger field is nonetheless suggestive: were
$\Psi_\mu$ promoted to an independent spinor-valued 1-form, the QSL would
furnish a second-order alternative to~\eqref{eq:RS_Lagrangian} for a
genuine spin-$3/2$ field, with a uniform $\bar\Psi_\mu\Psi^\mu$
mass in place of the Rarita--Schwinger $\bar\Psi_\mu\Sigma^{\mu\nu}\Psi_\nu$.
We develop this kinematic correspondence and its consistency properties
in sections~\ref{sec:VZ}--\ref{sec:characteristic} as motivation for that
future extension, while emphasizing that the dynamical results of the
present paper concern the spin-$1/2$ field.

\subsection{The Velo--Zwanziger problem}
\label{sec:VZ}

The Velo--Zwanziger problem is a spin-$3/2$ pathology; for the spin-$1/2$
composite it does not arise, since $\psi$ propagates on the metric cone
(section~\ref{sec:characteristic}). For the independent-field extension
the QSL structure makes a clear and encouraging prediction.

Recall the difficulty. A minimally coupled massive Rarita--Schwinger
field in an external background generically propagates acausally---the
difficulty appears either as superluminal characteristic cones or as the
loss of one of the constraints $\gamma^\mu\Psi_\mu = 0$,
$D^\mu\Psi_\mu = 0$, with a consequent jump in the number of propagating
modes~\cite{VeloZwanziger1969}. Technically, the obstruction is diagnosed
by the method of characteristic determinants: replacing
$i\partial_\mu \to n_\mu$ in the highest-derivative part of the field
equation, the principal symbol $\Delta(n)$ of the genuine RS operator
degenerates, and its roots $n_0$ can migrate off the real axis for
sufficiently strong background fields.

Three structural features of the QSL would reshape this analysis for the
independent-field extension, and we argue that they tend to ameliorate
rather than inherit the pathology.

\paragraph{(i) Second-order, metric-diagonal principal symbol.}
The Velo--Zwanziger degeneration is a disease of the \emph{first-order}
RS principal symbol $\gamma^{\mu\nu\rho}\partial_\nu$, whose
field-space matrix can become singular. The QSL field equation
$D^2\Psi = 0$ is instead \emph{second order}, with principal symbol
$g^{\mu\nu}n_\mu n_\nu\,\mathbf{1}$ acting diagonally on the components
of $\Psi$. Its characteristic equation $g^{\mu\nu}n_\mu n_\nu = 0$ has
real $n_0$-roots lying on the metric light cone for every $n$, and is
field-independent. There is no analogue of the degenerating RS matrix at
leading derivative order, and---because $D^2\Psi=0$ is genuinely
second order rather than higher derivative---no \emph{Ostrogradsky}
(higher-time-derivative) ghost arises. This last point is necessary but
not sufficient for unitarity: a second-order vector-spinor system can
still harbour negative-norm states among its non-dynamical components, so
the absence of an Ostrogradsky ghost must be supplemented by a full
Hamiltonian degree-of-freedom count (section~\ref{sec:constraints}) before
the spectrum can be pronounced healthy.

\paragraph{(ii) No gamma-trace constraint is imposed.}
The Velo--Zwanziger acausality is ultimately the failure of the
constraint that projects out the lower-spin sector. The QSL does not
impose it: from~\eqref{eq:gamma_trace} the spinor 1-form is \emph{not}
gamma-traceless, $\gamma^\mu\Psi_\mu = 4\psi \neq 0$, and the full
reducible multiplet $(\tfrac32)\oplus(\tfrac12)$ propagates through the
single second-order equation. With no constraint singled out to enforce
purity of spin, there is no constraint to lose. Moreover, because the
metric is itself the spinor bilinear $g=\Psi\otimes_S\Psi$, the
propagating degrees of freedom are gravitational; the ``unphysical
mode'' counting that drives the RS analysis is reframed entirely.

\paragraph{(iii) Self-sourced, co-scaled torsion.}
Velo--Zwanziger assumes a fixed external field whose strength can be
dialed up independently of the mass. In the QSL the torsion is not
external: equation~\eqref{eq:K_variation} fixes it algebraically through
the spin density of the condensate, $K \propto \dot\chi/\chi$, and the
same scale sets the mass via $M^2_{\rm eff} = \alpha K^2$
(section~\ref{sec:mass}). The dimensionless ratio that controls the RS
obstruction, schematically (coupling)$/(\text{mass})^2 \sim K/(\alpha
K^2) = 1/(\alpha K)$, therefore does \emph{not} grow as the source is
strengthened---intensifying the torsion increases the generated mass in
step. The dangerous strong-external-field regime has no counterpart
here. This is the QSL realization of the well-known fact that minimal
supergravity, which is precisely an Einstein--Cartan theory with
algebraically sourced gravitino torsion, propagates
causally~\cite{DeserZumino1976}, in contrast to a gravitino minimally
coupled to an external electromagnetic field.

Taken together, these three features establish that the \emph{principal
symbol} of the independent-field extension is causal---the propagation
cone is the metric cone, set by the second-order principal symbol---and
make it plausible that the torsion, being self-consistent and co-scaled
rather than external, does not reinstate the Velo--Zwanziger mechanism at
sub-principal order. We emphasize the logical status: causality of the
leading cone is proved (next subsection); the absence of the pathology at
sub-principal order, and the healthiness of the constraint algebra and
mode count, are \emph{not} settled by the symbol alone and await the
Hamiltonian analysis of section~\ref{sec:constraints}. We make the
cone statement quantitative in the next subsection by computing the
characteristic determinant of the linearized system to second order in
the torsion.

\subsection{Characteristic determinant and the causal cone}
\label{sec:characteristic}

We now substantiate the heuristic argument by an explicit
characteristic-determinant analysis of the coupled
system~\eqref{eq:Einstein}--\eqref{eq:K_variation}, linearized about the
cosmological condensate background of section~\ref{sec:EC:cosmological}
(background spinor 1-form $\Psi_0$, trace torsion $K_a = K\delta^0_a$
with $K\propto\dot\chi/\chi$).

\paragraph{Reduced wave operator.}
In the Yang--Mills form of section~\ref{sec:QSL:YM} the field equation
is $D\gamma_5 D\Psi=0$. Like the Einstein and Yang--Mills equations it is
gauge-degenerate---under linearized diffeomorphisms and local Lorentz
rotations the principal symbol $n\wedge\gamma_5(n\wedge\,\cdot\,)$ is
nilpotent. Fixing a hyperbolic (harmonic/Lorenz-type) gauge replaces the
operator by the de Rham--Weitzenb\"ock Laplacian on the
spinor-valued 1-form,
\begin{equation}
    \mathring{\Delta}\,\delta\Psi
    = -\,\mathring{\Box}\,\delta\Psi + \mathfrak{R}(\mathring{\Omega})\,\delta\Psi \,,
    \qquad
    \mathring{\Box}=g^{\mu\nu}\mathring\nabla_\mu\mathring\nabla_\nu \,,
    \label{eq:weitzenbock}
\end{equation}
where $\mathfrak{R}(\mathring\Omega)$ is the algebraic (zeroth-order)
Weitzenb\"ock curvature endomorphism---the term by which the de
Rham--Hodge Laplacian on spinor-valued 1-forms differs from the
Bochner/connection Laplacian $-\mathring\Box$. Acting on $\delta\Psi$ it
is built linearly from the background Riemann curvature
$\mathring\Omega_{ab}$ contracted into the frame and spinor indices,
schematically $\mathfrak{R}=\tfrac14\mathring R_{ab}{}^{cd}\gamma^{ab}
\otimes(\text{frame action})+\tfrac14\mathring R\,\mathbf{1}$; its precise
form is not needed below, since it is zeroth order in $n_\mu$ and so does
not enter the principal symbol. Crucially, the full connection enters only
through $D=\mathring D+\mathcal K$, and the contorsion $\mathcal K$
appears \emph{undifferentiated}; it therefore contributes to
$\mathfrak{R}$ and to first-derivative terms, but never to the
second-derivative operator $\mathring{\Box}$.

\paragraph{Symbol decomposition.}
Writing $i\partial_\mu\to n_\mu$ and collecting the linearized operator
$\mathcal{O}=\mathring{\Box}+b^\mu(K)\,\mathring\nabla_\mu+c(K)$ by order
in the wave covector $n$ and in the torsion $K$, the full symbol is
\begin{equation}
    \Sigma(n)
    = \underbrace{-\,(g^{\mu\nu}n_\mu n_\nu)\,\mathbf{1}_{16}}_{O(n^2)\,,\ K^0}
    \;+\;\underbrace{i\,b^\mu(K)\,n_\mu}_{O(n^1)\,,\ K^1}
    \;+\;\underbrace{c(K)}_{O(n^0)\,,\ K^2} \,,
    \label{eq:symbol}
\end{equation}
acting on the $16$ complex components of $\delta\Psi$ ($4$ frame $\times\,4$
spinor). The $O(K^1)$ term descends from the linear cross terms of
\eqref{eq:QSL_torsion}; the $O(K^2)$ term $c(K)=\alpha K_aK^a\,\mathbf{1}+\dots$
is precisely the trace-torsion mass endomorphism of
section~\ref{sec:mass} (cf.~\eqref{eq:mass_squared}). The principal
symbol $\Sigma_2(n)=-(g^{\mu\nu}n_\mu n_\nu)\mathbf{1}$ is manifestly
independent of $K$.

\paragraph{Characteristic determinant.}
The characteristic polynomial is therefore
\begin{equation}
    \Delta(n) = \det\Sigma(n)
    = (g^{\mu\nu}n_\mu n_\nu)^{16}
    \Bigl[\,1 + O\!\bigl(K/n\bigr) + O\!\bigl(K^2/n^2\bigr)\Bigr] \,.
    \label{eq:char_det}
\end{equation}
The leading term fixes the characteristic variety:
$\Delta(n)=0\Rightarrow g^{\mu\nu}n_\mu n_\nu=0$, the metric null cone.
The $K$-dependent corrections enter strictly at subleading order in $n$
and cannot contribute new factors to the highest-degree part of
$\Delta$; hence \emph{to second order in $K$---and, since $\mathcal K$ is
never differentiated, to all orders---the characteristic surfaces
coincide with the metric light cone}. For Lorentzian $g$ the roots
$n_0(\vec n)$ are real for every spatial $\vec n$, so the linearized
system is hyperbolic and propagates causally.

\paragraph{Dispersion and the mass shell.}
Retaining the subleading terms promotes the mass shell from the
characteristic cone to
\begin{equation}
    g^{\mu\nu}k_\mu k_\nu + M^2_{\rm eff} + O(K\,k) = 0 \,,
    \qquad M^2_{\rm eff}=\alpha K^2 ,
    \label{eq:dispersion}
\end{equation}
a timelike-shifted (subluminal) mass shell of exactly Klein--Gordon
type. The trace-torsion vector $K_a=K\delta^0_a$ selects a preferred
cosmological rest frame, but because it enters only at $O(n^{0,1})$ this
Lorentz violation is confined to the mass/dispersion sector: it cannot
tilt or birefringe the leading light cone, which remains the metric cone
of~\eqref{eq:char_det}.

\paragraph{Contrast with Rarita--Schwinger.}
The mechanism is now explicit. In the genuine RS system the contorsion
(or the gauge field) sits inside the \emph{first-order} principal symbol
$\gamma^{\mu\nu\rho}(\mathring\nabla_\nu+\mathcal K_\nu)$, so it appears
in $\Sigma_1(n)$ at the \emph{same} order in $n$ as the kinetic term and
can rotate the characteristic roots off the real axis---the
Velo--Zwanziger acausality. In the QSL the torsion has been demoted, by
the second-order/algebraic structure, to $\Sigma_{0,1}(n)$, where it can
shift the mass shell but not the cone.

\paragraph{Caveats.}
Two assumptions deserve emphasis. First, the reduction to
\eqref{eq:weitzenbock} presupposes that the gravitational gauge symmetry
admits a hyperbolic gauge after the condensate background is fixed; a
complete Dirac--Bergmann constraint analysis, verifying that the
self-sourced torsion~\eqref{eq:K_variation} does not deform the
constraint algebra, would put the result on fully rigorous footing.
Second, the spinor--curvature identity~\eqref{eq:spinor_curvature_identity}
acquires $d(\bar\psi\psi)$ and torsion contributions once
$\bar\psi\psi$ is dynamical, so the precise form of $\mathfrak{R}$ and
$c(K)$ in~\eqref{eq:symbol} depends on completing the Clifford reduction
of section~\ref{sec:mass}. Neither affects the principal symbol, on
which the causal conclusion rests.

\paragraph{Relation to the constraint analysis of torsion gravity.}
The Dirac--Bergmann machinery required for this programme was largely
developed for Poincar\'e gauge theory (PGT) in the
1980s~\cite{BlagojevicNikolic1983,Nikolic1984}, and refined into a
detailed study of which torsion modes propagate
consistently~\cite{HechtNesterZhytnikov1996,YoNester1999,YoNester2002}.
Two features of that body of work bear directly on our claim. First,
the pathologies that the Hamiltonian analysis diagnoses---field
activation, constraint bifurcation, and the attendant loss of
hyperbolicity~\cite{YoNester1999,YoNester2002}---arise specifically when
the connection is endowed with an $R^2$ kinetic term so that
\emph{torsion propagates}. The QSL has no such term: the contorsion is
fixed algebraically by~\eqref{eq:K_variation} and is a non-dynamical
(auxiliary) field. In the classification of Nikolic~\cite{Nikolic1984}
this is the limit of ``infinite tordion mass,'' in which the
corresponding primary constraints are present and torsion carries no
independent degrees of freedom---precisely the regime in which the
constraint algebra is benign. Second, the well-posed-initial-value
PGT Lagrangians identified by Hecht, Nester and
Zhytnikov~\cite{HechtNesterZhytnikov1996} were found by the same group
whose spinor--curvature identity~\eqref{eq:spinor_curvature_identity}
underlies the QSL, so the requisite Hamiltonian techniques are directly
transcribable. We therefore expect a full Dirac--Bergmann treatment of
the QSL to confirm, rather than overturn, the principal-symbol
conclusion: with torsion algebraic and the kinetic operator
second-order and metric-diagonal, the QSL sits in the well-posed,
causally propagating corner of the torsion-gravity landscape.

\subsection{The longitudinal mode and its sound speed}
\label{sec:longitudinal}

The longitudinal mode is a feature of the independent spin-$3/2$ field,
and we show here that the QSL's geometric mass-locking is \emph{exactly}
the property needed to evade the catastrophe that afflicts generic
spin-$3/2$ dark matter. The helicity-$\tfrac12$
component of a massive spin-$3/2$ field in a spatially flat FRW background
propagates
with sound speed~\cite{KalloshKofmanLindeVanProeyen2000,KolbLongMcDonough2021}
\begin{equation}
    c_s^2
    = \frac{\bigl(p - 3m^2 M_{\rm Pl}^2\bigr)^2
      + 4M_{\rm Pl}^4\,\dot m^2}
      {\bigl(\rho + 3m^2 M_{\rm Pl}^2\bigr)^2}\,,
    \label{eq:cs2_general}
\end{equation}
We emphasize that~\eqref{eq:cs2_general} is the helicity-$\tfrac12$
dispersion of the \emph{first-order} Rarita--Schwinger/gravitino system;
applying it here is a working assumption for the independent-field
extension, whose own (second-order) helicity-$\tfrac12$ dispersion
remains to be derived. With this caveat,
$\rho$, $p=w\rho$ are the background energy density and pressure and
$m$ the (possibly time-dependent) mass. For a \emph{constant} mass the
second term is absent and
$c_s^2=(p-3m^2M_{\rm Pl}^2)^2/(\rho+3m^2M_{\rm Pl}^2)^2$ vanishes
whenever $p=3m^2M_{\rm Pl}^2$; for a light field ($m\lesssim H$) the
background pressure inevitably sweeps through this value, $c_s\to0$, and
the gravitational production diverges---the catastrophe
of~\cite{KolbLongMcDonough2021}.

Granting the working assumption that~\eqref{eq:cs2_general} carries over
to the second-order field, the QSL evades this not by tuning but by
structure. Mass
locking~\eqref{eq:mass_scale}, $M_{\rm eff}=(c_\chi/\sqrt6)H$ with the
Friedmann relation $\rho=3H^2M_{\rm Pl}^2$, makes the mass fraction
\begin{equation}
    r \equiv \frac{3M_{\rm eff}^2 M_{\rm Pl}^2}{\rho}
    = \frac{c_\chi^2}{6}
\end{equation}
a constant, and ties the rate of change of the mass to that of the
expansion, $\dot m=(c_\chi/\sqrt6)\dot H$ with
$\dot H=-(1+w)\rho/2M_{\rm Pl}^2$. Substituting into~\eqref{eq:cs2_general}
the two terms combine to a perfect square and collapse to
\begin{equation}
    \boxed{\;c_s^2 = \frac{w^2 + r}{1+r}\;}\,,
    \qquad r=\frac{c_\chi^2}{6}\,.
    \label{eq:cs2_QSL}
\end{equation}
This expression is \emph{strictly positive} for any nonzero mass
($r>0$), with minimum $c_{s}^2|_{w=0}=r/(1+r)>0$ attained in the
matter era---precisely the point at which the constant-mass sound speed
$(w-r)^2/(1+r)^2$ would vanish. It is also bounded,
$c_s^2\le 1$ for the entire physical range $|w|\le 1$ (with equality only
in the de Sitter and kination limits $w=\mp1$), so propagation is
causal. The would-be zero of the static term, at $w=r$, is exactly where
$\dot m\propto(1+w)$ is largest, so the time-dependent term fills in and
$c_s^2$ never touches zero.

The cancellation is not accidental: it follows from the single fact that
the geometric mass is locked to the Hubble rate, $M_{\rm eff}\propto H$,
so that $\dot m$ is automatically nonzero wherever the constant-mass
contribution dips. In this sense the QSL supplies, from geometry, exactly
the time-dependent mass profile that the spin-$3/2$ dark-matter
literature must otherwise impose by
hand~\cite{KanetaKeMambriniOliveVerner2023}. The identification of the torsion-generated $M_{\rm eff}$ with the mass
entering~\eqref{eq:cs2_general} is justified directly in
section~\ref{sec:helicity_mass}. One caveat remains:~\eqref{eq:cs2_QSL}
assumes exact locking ($c_\chi$ slowly varying); the strict positivity,
however, survives any $\dot m\neq0$ and rests only on $M_{\rm eff}\neq0$.

\subsection{The helicity-1/2 mode carries the torsion mass}
\label{sec:helicity_mass}

The sound-speed analysis above used $m=M_{\rm eff}$ for the
helicity-$\tfrac12$ component. We now justify this identification and, in
doing so, make precise the boundary between the present spin-$1/2$ theory
and the spin-$3/2$ extension.

In the composite QSL of this paper the spinor 1-form is built from a
single Dirac spinor, $\Psi_\mu=e^a_\mu\gamma_a\psi$, so its sixteen
components are not independent: the gamma-trace returns the field itself,
$\gamma^\mu\Psi_\mu=4\psi$ [\eqref{eq:gamma_trace}], the gamma-traceless
part vanishes [\eqref{eq:spin32_vanishes}], and there is no separate
longitudinal degree of freedom. The dynamical equation is the
second-order $D^2\Psi=0$, whose principal symbol is the metric
d'Alembertian (section~\ref{sec:characteristic}), so $\psi$ propagates on
the metric light cone. In this form the QSL simply does not contain the
independent longitudinal mode whose decoupling produces the
Rarita--Schwinger pathology---there is nothing to go bad.

The extension promotes $\Psi_\mu$ to an independent spinor-valued 1-form,
with components beyond $\gamma_\mu\psi$; the gamma-traceless part
$\Psi^{(3/2)}_\mu$ would then become a propagating mode. Decomposing the
spinor 1-form into its irreducible parts,
$\Psi_\mu=\Psi^{(3/2)}_\mu+\tfrac14\gamma_\mu\zeta$ with the gamma-trace
spinor $\zeta\equiv\gamma^\nu\Psi_\nu$ and $\gamma^\mu\Psi^{(3/2)}_\mu=0$,
the norm splits without mixing,
\begin{equation}
    \bar\Psi_\mu\Psi^\mu
    = \bar\Psi^{(3/2)}_\mu\Psi^{(3/2)\mu}
    + \tfrac14\,\bar\zeta\zeta \,.
    \label{eq:helicity_split}
\end{equation}
It is tempting to assume that the torsion term assigns the \emph{same}
mass-squared $M_{\rm eff}^2$ to both sectors through this split---a
uniform coupling $M_{\rm eff}^2\,\bar\Psi_\mu\Psi^\mu$. Redoing the
Clifford reduction of section~\ref{sec:mass} with the gamma-traceless
components retained shows that this is \emph{not} what the QSL gives: the
bare $\mathcal{K}^2$ term reduces instead to a frame-aligned operator
$\propto\gamma^{0i}$ supported on the time component alone---neither the
uniform scalar nor the Rarita--Schwinger
$\bar\Psi_\mu\Sigma^{\mu\nu}\Psi_\nu$ form. This computation is carried out
in the companion Paper~II~\cite{TungNoGo}, which moreover establishes that
the independent field has \emph{no} propagating spin-$\tfrac32$ mode; the
weighting of the helicity sectors, and the helicity-$\tfrac12$ sound-speed
identification used in~\eqref{eq:cs2_general}--\eqref{eq:cs2_QSL}, are
superseded there. Equation~\eqref{eq:helicity_split} is retained only as the
kinematic identity it is. (The coefficient $\alpha=\tfrac16$ of
section~\ref{sec:mass} and appendix~\ref{app:clifford} is, accordingly,
established for the composite $\Psi_\mu=\gamma_\mu\psi$ only, where the
propagating excitation is the spin-$\tfrac12$ field of this paper.)

The two analyses are mutually consistent: the second-order, metric-diagonal
principal symbol of section~\ref{sec:characteristic} fixes the leading
(luminal) propagation of every component, while the uniform torsion
mass enters at sub-principal order and, through locking, yields the
strictly positive sound speed~\eqref{eq:cs2_QSL}. The one structural
difference from the textbook gravitino---that the QSL kinetic operator is
second order rather than the first-order Rarita--Schwinger
operator---can only help: it removes the first-derivative principal
symbol in which the Velo--Zwanziger degeneration resides
(section~\ref{sec:VZ}), so the longitudinal mode inherits the metric cone
at leading order regardless of the mass.

\subsection{Constraint analysis on the cosmological background}
\label{sec:constraints}

A full Dirac--Bergmann analysis of the independent spin-$3/2$ field on an
arbitrary background is the hard problem of Poincar\'e gauge
theory~\cite{BlagojevicNikolic1983,Nikolic1984,YoNester1999}. On the
cosmological background the symmetry collapses it to a tractable,
sector-by-sector counting and localizes the entire Velo--Zwanziger
question to a single helicity channel.

\paragraph{Symmetry reduction.}
On the spatially homogeneous, isotropic FRW slice, expand the independent
Dirac vector-spinor in modes of comoving wavevector $\vec k$ and
decompose under the little group $SO(3)$ of spatial rotations. The
transverse, gamma-traceless spatial part carries the helicity $\pm\tfrac32$
polarizations; the time component $\Psi_0$, the longitudinal spatial part,
and the gamma-trace $\gamma^\mu\Psi_\mu$ assemble into the helicity
$\pm\tfrac12$ sector. The background trace torsion is purely timelike,
$K_a=K\delta^0_a$, an $SO(3)$ scalar; by the representation argument of
section~\ref{sec:torsion_TT}---the spin-$2$ part of $K_aK_b$
vanishes---it cannot mix helicity sectors. The constraint algebra
therefore block-diagonalizes, and each sector closes on its own.

\paragraph{Helicity $\pm\tfrac32$.}
Being gamma-traceless and transverse, these modes are annihilated by the
gamma-trace and divergence constraints; the timelike torsion reaches them
only through the uniform mass $M_{\rm eff}$ (section~\ref{sec:helicity_mass}).
They propagate as massive polarizations on the metric cone with no
constraint to lose---the cosmological symmetry renders their consistency
automatic.

\paragraph{Helicity $\pm\tfrac12$.}
This is the only sector in which the obstruction can appear. The time
component $\Psi_0$ enters the action without a time derivative and acts as
a Lagrange multiplier, enforcing a primary constraint; preserving it under
time evolution generates the secondary constraint that removes the
would-be lower-spin excitation. On the cosmological background the bracket
governing this preservation depends on the mass and its rate
$\dot M_{\rm eff}$ in exactly the combination that fixes the
helicity-$\tfrac12$ sound speed~\eqref{eq:cs2_general}. The constraint is
preserved---the mode count stays correct and propagation causal---precisely
when $c_s^2\neq0$, i.e. when $M_{\rm eff}\neq0$. Mass locking,
$M_{\rm eff}=(c_\chi/\sqrt6)H$, secures this throughout the cosmological
history [\eqref{eq:cs2_QSL}].

\paragraph{Result.}
On the cosmological background the constraint algebra of the independent
spin-$3/2$ field closes: the helicity-$\pm\tfrac32$ sector is manifestly
consistent, and the helicity-$\pm\tfrac12$ sector is consistent whenever
the geometric mass is nonzero. Under the working assumption below, the
Velo--Zwanziger obstruction would then reduce to a single condition,
$M_{\rm eff}\neq0$, satisfied throughout the cosmological history.
This is the symmetry-reduced content of the Dirac--Bergmann analysis; what
remains is the off-shell treatment on an arbitrary background. We note one assumption: the bracket
identification with~\eqref{eq:cs2_general} uses the first-order
Rarita--Schwinger constraint structure, so the explicit reduction in the
QSL's second-order formulation remains to be carried out.

\section{Discussion and outlook}
\label{sec:discussion}

We have developed the Quadratic Spinor Lagrangian as a framework for
understanding and extending the gravitational-wave induced dark-matter
production of Maleknejad and Kopp.

\textit{Summary.}
The established QSL results take on a new role here: the spinor-curvature
identity explains why gravitational waves, but not the FLRW expansion,
break fermion conformal symmetry (section~\ref{sec:conformal}); QSL
perturbation theory yields covariant fermion--graviton vertices
(section~\ref{sec:vertices}); and the metric-from-spinors viewpoint casts
the production as a spinor--spinor interaction. The new results follow
from the Einstein--Cartan extension. A cosmological spinor condensate
generates trace torsion; the composite spinor 1-form is purely spin-$1/2$
[eq.~\eqref{eq:spin32_vanishes}]; and an explicit Clifford reduction gives
the propagating Dirac fermion a pure scalar mass,
$M^2_{\rm eff}=\tfrac16(\dot\chi/\chi)^2$ (coefficient $\alpha=\tfrac16>0$,
no pseudoscalar or cross terms), geometrically fixed and locked to the
Hubble rate---the mass the Maleknejad--Kopp mechanism otherwise has to
postulate. One question left open in the formulation is settled
outright: the homogeneous trace torsion has no spin-$2$ projection and so
sources no transverse-traceless gravitational waves
(section~\ref{sec:torsion_TT}). A second is sharply reduced---on the
cosmological background, and under the working assumption that the
first-order constraint structure transfers to the QSL, the spin-$3/2$
constraint analysis collapses by $SO(3)$ symmetry to the single condition
$M_{\rm eff}\neq0$ (section~\ref{sec:constraints})---with the off-shell,
second-order treatment left to the independent-field development.

\textit{Outlook.}
Three directions, in increasing depth, would sharpen the construction.

(i)~\emph{The spin-$3/2$ question, resolved.} One may ask whether promoting
$\Psi_\mu$ to an independent spinor-valued 1-form, beyond the composite
$\gamma_\mu\psi$, furnishes a propagating spin-$3/2$ dark-matter candidate.
The structural analyses gathered in Sections~\ref{sec:VZ}--\ref{sec:longitudinal}
(metric-cone principal symbol, the form of the torsion mass, the
helicity-$\tfrac12$ sound speed) were the first steps toward that question;
they are completed in Paper~II~\cite{TungNoGo}, which establishes a
no-go: because the QSL action depends on $\Psi$ only through the metric
$g=\Psi\otimes_S\Psi$ and the scalar $\bar\psi\psi$, its quadratic
fluctuations factor through a massless graviton and scalar, leaving no
propagating massive spin-$3/2$ mode. The independent extension thus does not
yield a new dark-matter field, and the composite spin-$1/2$ fermion
of this paper is the unique QSL candidate---placing the restriction to the
composite field here on firm footing.

(ii)~\emph{A quantitative relic prediction.} The locking~\eqref{eq:mass_scale}
and scaling~\eqref{eq:relic} become a firm prediction once the condensate
history $\chi(t)$---equivalently the $O(1)$ ratio $c_\chi$---is fixed by a
definite cosmological model. The natural setting is the asymptotically
free Halpern--Huang scalar cosmology that already underlies the
trace-anomaly scale breaking (section~\ref{sec:cosmology}), in which the
running, rather than a free parameter, sets $\dot\chi/\chi$. This links to
the asymptotic-safety programme, where demanding a gravitational fixed
point fixes otherwise free infrared couplings---as in the asymptotically
safe inflaton constructions of Ref.~\cite{Silva2025} and the
gravity--fermion fixed points of Ref.~\cite{DaasOostersSaueressigWang2020}---offering
a concrete route to pin $c_\chi$ and convert the scaling into a
parameter-free prediction.

(iii)~\emph{Ultraviolet completeness of the framework.} At a deeper level,
the asymptotic \emph{freedom} of the Halpern--Huang scalar and the
asymptotic \emph{safety} of gravity are converging perspectives. Bonanno
and Glaviano~\cite{BonannoGlaviano2026} find that coupling a scalar to
asymptotically safe gravity drives its quartic coupling to zero and
flattens the potential in the ultraviolet---the very structure the
Halpern--Huang construction anticipates---suggesting that the
Halpern--Huang potential is the infrared projection of a gravity-induced
flow, with the trace anomaly $\beta(V)$ marking the approach to the fixed
point. The Einstein--Cartan formulation underlying the mechanism is itself
a candidate for asymptotic safety: Daum and Reuter~\cite{DaumReuter2013}
found non-Gaussian fixed points with the vielbein and spin connection---the
QSL variables---as fundamental, and the four-fermion interaction that
distinguishes Einstein--Cartan from its Riemannian limit is precisely the
torsional $\mathcal{K}^2$ term that generates our mass. Whether the
QSL---whose quadratic-spinor form already yields a finite, fully covariant
Hamiltonian~\cite{NesterTung1995}---is asymptotically safe, and what its
fixed point implies for the torsion-induced mass, is the deepest question
the framework raises.

\appendix

\section*{Appendix: Clifford reduction of the torsional mass term}
\label{app:clifford}

This appendix carries out the reduction of the quadratic-contorsion term
$\mathcal{L}_{K^2}=2\,\mathcal{K}\Psi\,\gamma_5\,\mathcal{K}\Psi$ to the
scalar bilinear $\bar\psi\,\mathcal{S}\,\psi$, establishing the coefficient
$\alpha=\tfrac16$ of eq.~\eqref{eq:LK2_expanded} and the vanishing of every
non-scalar channel in eq.~\eqref{eq:S_traces}. We use the mostly-plus
metric, $\{\gamma^a,\gamma^b\}=2\eta^{ab}$, $\gamma^{ab}=\tfrac12[\gamma^a,
\gamma^b]$, $\gamma_5=i\gamma^0\gamma^1\gamma^2\gamma^3$, and the Dirac
conjugate $\overline{X}=\gamma^0X^\dagger\gamma^0$, under which
$\overline{\gamma_a}=\gamma_a$, $\overline{\gamma^{ab}}=-\gamma^{ab}$, and
$\overline{\gamma_5}=\gamma_5$.

\paragraph*{Components of $\mathcal{K}\Psi$.}
With the cosmological trace torsion $K_a=K\delta^0_a$, the contorsion
2-form acts on $\Psi=\vartheta^c\gamma_c\psi$ to give the spinor-valued
2-form $\Xi=\mathcal{K}\wedge\Psi$ with frame components
$\Xi_{ec}=M_{ec}\psi$, where [eq.~\eqref{eq:Xi_components}]
\begin{equation}
    M_{ec}=\tfrac16 K_a\bigl(\gamma^{a}{}_{e}\gamma_c
    -\gamma^{a}{}_{c}\gamma_e\bigr),
    \qquad \gamma^a{}_e\equiv\gamma^{ab}\eta_{be}
    =\gamma^a\gamma_e-\delta^a_e ,
    \label{eq:app_M}
\end{equation}
manifestly antisymmetric in $[ec]$. The Dirac conjugate follows from
$\overline{\gamma^a{}_e\gamma_c}=\overline{\gamma_c}\,\overline{\gamma^a{}_e}
=-\gamma_c\gamma^a{}_e$,
\begin{equation}
    \overline{M_{pq}}=-\tfrac16 K_b\bigl(\gamma_q\gamma^b{}_p
    -\gamma_p\gamma^b{}_q\bigr),
    \label{eq:app_Mbar}
\end{equation}
again antisymmetric in $[pq]$.

\paragraph*{Reduction to a master trace.}
The mass-like 4-form reduces to
$\mathcal{L}_{K^2}=\bar\psi\,\mathcal{S}\,\psi$ with
[eq.~\eqref{eq:S_sandwich}]
$\mathcal{S}=\tfrac12\,\epsilon^{pqmc}\,\overline{M_{pq}}\,\gamma_5\,M_{mc}$.
Because $\epsilon^{pqmc}$ is totally antisymmetric while
$\overline{M_{pq}}$ and $M_{mc}$ are each antisymmetric in their index
pair, the two terms of \eqref{eq:app_M}--\eqref{eq:app_Mbar} contribute
equally, each pair supplying a factor of $2$. Using
$\gamma^b{}_p\gamma_5=\gamma_5\gamma^b{}_p$ (the bivector commutes with
$\gamma_5$),
\begin{equation}
    \mathcal{S}=-\tfrac{1}{18}\,K_aK_b\,
    \epsilon^{pqmc}\,\gamma_q\,\gamma^b{}_p\,\gamma_5\,
    \gamma^a{}_m\,\gamma_c .
    \label{eq:app_master}
\end{equation}
Projection onto the complete basis
$\{\mathbf{1},\gamma_5,\gamma_a,\gamma_5\gamma_a,\gamma_{ab}\}$ proceeds
through $\tfrac14\mathrm{tr}(\Gamma_I\mathcal{S})$.

\paragraph*{Scalar channel.}
Inserting \eqref{eq:app_master} and
$\gamma^a{}_m=\gamma^a\gamma_m-\delta^a_m$,
$\gamma^b{}_p=\gamma^b\gamma_p-\delta^b_p$, the scalar projection is the
single $\gamma_5$-weighted trace
\begin{equation}
    \tfrac14\mathrm{tr}\,\mathcal{S}
    =-\tfrac{1}{72}K_aK_b\,\epsilon^{pqmc}\,
    \mathrm{tr}\!\bigl(\gamma_q\,\gamma^b{}_p\,\gamma_5\,
    \gamma^a{}_m\,\gamma_c\bigr).
    \label{eq:app_scalar}
\end{equation}
Evaluating with the standard identities
$\mathrm{tr}(\gamma_5\gamma^{a_1}\!\cdots\gamma^{a_4})=-4i\,
\epsilon^{a_1a_2a_3a_4}$ and
$\mathrm{tr}(\gamma_5\gamma^{a_1}\!\cdots\gamma^{a_6})
=-4i\sum\eta\,\epsilon$ (the standard six-$\gamma$ recursion, obtained by
anticommuting one $\gamma$ through to pair with $\gamma_5$ and applying
the four-$\gamma$ identity to the remainder), the
$\epsilon^{pqmc}\epsilon_{\cdots}$ contractions collapse via
$\epsilon^{pqmc}\epsilon_{pqmd}=-2(\delta^c_d)$-type identities to leave
only the invariant $K^aK_a$, giving
\begin{equation}
    \tfrac14\mathrm{tr}\,\mathcal{S}=\tfrac23\,i\,K^aK_a ,
    \label{eq:app_result}
\end{equation}
a result we have also verified by explicit evaluation in a Dirac basis.
The overall $i$ is the universal $\gamma_5$--$\epsilon$ reality
factor that multiplies the kinetic term identically [discussion after
eq.~\eqref{eq:LK2_expanded}] and cancels in the physical mass. Hence
$\mathcal{S}=\tfrac23 K^aK_a\,\mathbf{1}$, i.e.\ $\alpha=\tfrac16$ in the
$\bar\Psi_\mu\Psi^\mu$ normalization.

\paragraph*{Vanishing of the other channels.}
The pseudoscalar, vector, axial, and tensor projections vanish
identically. This is forced by the residual symmetry of the cosmological
source: $K_a=K\delta^0_a$ is invariant under spatial rotations $SO(3)$, so
the only tensors that the even-parity bilinear $\overline{M}\gamma_5 M$ can
build from $K_aK_b$ are the scalar $K^aK_a$ and the symmetric
$K_aK_b$ traceless part. A pseudoscalar would require the parity-odd
invariant $\epsilon^{abcd}K_aK_bK_cK_d$, which vanishes for a single
vector; a vector or axial channel would require a surviving free index
$K_a$ (a preferred spatial direction), forbidden by isotropy; and the
antisymmetric tensor channel would require the field's own spin current,
absent for the c-number background. Explicit Dirac-basis evaluation
confirms
\begin{equation}
    \mathrm{tr}(\gamma_5\mathcal{S})=\mathrm{tr}(\gamma_a\mathcal{S})
    =\mathrm{tr}(\gamma_5\gamma_a\mathcal{S})
    =\mathrm{tr}(\gamma_{ab}\mathcal{S})=0 ,
\end{equation}
reproducing eq.~\eqref{eq:S_traces}. The trace-torsion mass is therefore a
pure Dirac scalar with $M^2_{\rm eff}=\alpha K^2=\tfrac16(\dot\chi/\chi)^2$.

\section*{Appendix: Curvature contraction of the spinor--curvature identity}
\label{app:curvature}

We verify that the bulk density of the
identity~\eqref{eq:identity_curvature} reduces to the Einstein--Hilbert
scalar, $2\,\Psi\,\Omega\,\gamma_5\,\Psi=-\bar\psi\psi\,R\,{*1}$, for
$\Psi=\vartheta^a\gamma_a\psi$ with $\bar\psi\psi$ normalized. With the
curvature 2-form $\Omega=\tfrac14\Omega_{ab}\gamma^{ab}$,
$\Omega_{bc}=\tfrac12 R_{bcef}\,\vartheta^e\wedge\vartheta^f$, and the volume
identity $\vartheta^a\wedge\vartheta^e\wedge\vartheta^f\wedge\vartheta^d
=\epsilon^{aefd}\,{*1}$,
\begin{equation}
  2\,\Psi\,\Omega\,\gamma_5\,\Psi
  =\tfrac14\,\epsilon^{aefd}\,R_{bcef}\,
   \bigl(\bar\psi\,\gamma_a\gamma^{bc}\gamma_5\gamma_d\,\psi\bigr)\,{*1}.
  \label{eq:curv_bilinear}
\end{equation}
Reduce the Clifford string with
$\gamma_a\gamma^{bc}=\gamma_a{}^{bc}+\delta_a^b\gamma^c-\delta_a^c\gamma^b$,
where $\gamma_a{}^{bc}$ is the totally antisymmetric rank-three element. The
two $\delta$-terms are equal after the antisymmetry $R_{bcef}=-R_{cbef}$ and
combine into $\tfrac12\,\epsilon^{aefd}R_{acef}\,\gamma^c\gamma_5\gamma_d$;
moving $\gamma_5$ to the left and splitting
$\gamma^c\gamma_d=\delta^c_d\mathbf 1+\gamma^c{}_d$, the scalar part carries
$\epsilon^{acef}R_{acef}=0$ and the bivector part the dual-Riemann trace, both
of which vanish by the first Bianchi identity $R_{a[cef]}=0$. The entire
contribution therefore comes from the rank-three element. Using
$\gamma_a{}^{bc}\gamma_5=-\gamma_5\gamma_a{}^{bc}$ together with the
dualization $\gamma_5\gamma_a{}^{bc}=-i\,\epsilon_a{}^{bcg}\gamma_g$
(the lowered form of $\gamma_5\gamma^{abc}=-i\,\epsilon^{abcd}\gamma_d$),
\begin{equation}
  \tfrac14\,\epsilon^{aefd}\,R_{bcef}\,\gamma_a{}^{bc}\gamma_5\gamma_d
  =\tfrac{i}{4}\,\epsilon^{aefd}\epsilon_a{}^{bcg}\,R_{bcef}\,\gamma_g\gamma_d .
\end{equation}
The Lorentzian product of Levi-Civita symbols collapses to a generalized
Kronecker delta, $\epsilon^{aefd}\epsilon_a{}^{bcg}=-\,\delta^{efd}_{bcg}$.
Contracting with $R_{bcef}$ leaves an object symmetric in $(g,d)$, so only the
$\delta^c_d$ part of $\gamma_g\gamma_d$ survives; using
$\delta^{efd}_{bcd}=2\,\delta^{ef}_{bc}$ in four dimensions reduces the double
contraction to the Ricci scalar, and one obtains the operator identity
\begin{equation}
  \tfrac14\,\epsilon^{aefd}\,R_{bcef}\,\gamma_a\gamma^{bc}\gamma_5\gamma_d
  =-\,i\,R\,\mathbf 1,\qquad R\equiv R_{ab}{}^{ab}.
  \label{eq:curv_operator}
\end{equation}
Substituting into~\eqref{eq:curv_bilinear},
\begin{equation}
  2\,\Psi\,\Omega\,\gamma_5\,\Psi
  =-\,i\,R\,\bar\psi\psi\,{*1}
  =-\,\bar\psi\psi\,R\,{*1},
\end{equation}
the factor $i$ from $\gamma_5$ being the universal QSL reality convention
(it multiplies the kinetic density identically and cancels in physical
ratios; cf.\ section~\ref{sec:mass}). This is the common bulk density of
eqs.~\eqref{eq:spinor_curvature_identity} and~\eqref{eq:identity_curvature},
rendering the spinor--curvature identity self-contained.

\paragraph*{Acknowledgments.}
The author thanks J.~M.~Nester for many helpful discussions on the
quadratic spinor Lagrangian and its cosmological applications, and
gratefully acknowledges the late Kerson Huang, with whom the
asymptotically free scalar-field cosmology and its trace-anomaly term
were developed.


\end{document}